\documentclass[twocolumn]{aastex62}

\usepackage{verbatim}

\received{March 05, 2019}
\revised{February 21, 2020}
\accepted{March 04, 2020}

\shorttitle{\textit{RadioAstron} observations of S5~0716$+$71}
\shortauthors{Kravchenko et al.}
\begin{document}

\footnotesize\textsc{Accepted to Astrophysical journal}\vspace{1pt} \normalsize

\title{Probing the innermost regions of AGN jets and their magnetic fields with \textit{RadioAstron}.\\ III. Blazar S5~0716$+$71 at microarcsecond resolution}

\correspondingauthor{Evgeniya Kravchenko}
\email{e.kravchenko@ira.inaf.it}

\author[0000-0002-0786-7307]{E. V. Kravchenko}
\affil{INAF Istituto di Radioastronomia, Via P. Gobetti, 101, Bologna, 40129, Italy}
\affiliation{Astro Space Center, Lebedev Physical Institute, Russian Academy of Sciences, 
Profsoyuznaya st., 84/32, Moscow, 117997, Russia}

\author{J. L. G\'omez}
\affiliation{Instituto de Astrof\'isica de Andaluc\'ia, CSIC, Glorieta de la Astronom\'ia s/n, Granada, 18008, Spain}

\author{Y. Y. Kovalev}
\affiliation{Astro Space Center, Lebedev Physical Institute, Russian Academy of Sciences, 
Profsoyuznaya st., 84/32, Moscow, 117997, Russia}
\affiliation{Moscow Institute of Physics and Technology, Institutsky per. 9, Moscow region, Dolgoprudny, 141700, Russia}
\affiliation{Max-Planck-Institut f\"{u}r Radioastronomie, Auf dem H\"{u}gel 69, Bonn, 53121, Germany}

\author{A. P. Lobanov}
\affiliation{Max-Planck-Institut f\"{u}r Radioastronomie, Auf dem H\"{u}gel 69, Bonn, 53121, Germany}
\affiliation{Institut f\"{u}r Experimentalphysik, Universit\"{a}t Hamburg, Luruper Chaussee, 149, 22 761 Hamburg, Germany}

\author{T. Savolainen}
\affiliation{Aalto University Department of Electronic and Nanoengineering, PL15500, FI-00076 Aalto, Finland}
\affiliation{Aalto University Mets\"{a}hovi Radio Observatory, Mets\"{a}hovintie 114, FI-02540 Kylm\"{a}l\"{a}, Finland}
\affiliation{Max-Planck-Institut f\"{u}r Radioastronomie, Auf dem H\"{u}gel 69, Bonn, 53121, Germany}

\author{G. Bruni}
\affiliation{INAF-Istituto di Astrofisica e Planetologia Spaziali, via Fosso del Cavaliere, 100, 00133 Rome, Italy}

\author{A. Fuentes}
\affiliation{Instituto de Astrof\'isica de Andaluc\'ia, CSIC, Glorieta de la Astronom\'ia s/n, Granada, 18008, Spain}

\author{J. M. Anderson}
\affiliation{Deutsches GeoForschungsZentrum GFZ, Telegrafenberg, 14473 Potsdam, Germany}

\author{S. G. Jorstad}
\affiliation{Institute for Astrophysical Research, Boston University, 725 Commonwealth Avenue, Boston, MA 02215, USA}
\affiliation{Astronomical Institute, St. Petersburg University, Universitetskij Pr. 28, Petrodvorets, 198504 St. Petersburg, Russia}

\author{A. P. Marscher}
\affiliation{Institute for Astrophysical Research, Boston University, 725 Commonwealth Avenue, Boston, MA 02215, USA}

\author{M. Tornikoski}
\affiliation{Aalto University Mets\"{a}hovi Radio Observatory, Mets\"{a}hovintie 114, FI-02540 Kylm\"{a}l\"{a}, Finland}

\author{A. L\"{a}hteenm\"{a}ki}
\affiliation{Aalto University Mets\"{a}hovi Radio Observatory, Mets\"{a}hovintie 114, FI-02540 Kylm\"{a}l\"{a}, Finland}
\affiliation{Aalto University Department of Electronic and Nanoengineering, PL15500, FI-00076 Aalto, Finland}

\author{M. M. Lisakov}
\affiliation{Max-Planck-Institut f\"{u}r Radioastronomie, Auf dem H\"{u}gel 69, Bonn, 53121, Germany}
\affiliation{Astro Space Center, Lebedev Physical Institute, Russian Academy of Sciences, 
Profsoyuznaya st., 84/32, Moscow, 117997, Russia}

\begin{abstract}
We present \textit{RadioAstron} Space VLBI imaging observations of the BL~Lac object S5~0716$+$71 made on January 3--4 2015 at a frequency of 22\,GHz (wavelength $\lambda=1.3$~cm).
The observations were made in the framework of the AGN Polarization Key Science Program.
The source was detected on projected space-ground baselines up to 70\,833\,km (5.6 Earth diameters) for both, parallel hand and cross-hand interferometric visibilities.
We have used these detections to obtain a full-polarimetric image of the blazar at an unprecedented angular resolution of 24~$\mu$as, the highest for this source to date.
This enabled us to estimate the size of the radio core to be $<12\times5~\mu$as and to reveal a complex structure and a significant curvature of the blazar jet in the inner 100~$\mu$as, which is an indication that the jet viewing angle lies inside the opening angle of the jet conical outflow.
Fairly highly (15\%) linearly polarized emission is detected in a jet region of 19\,$\mu$as in size, located 58\,$\mu$as downstream from the core.
The highest brightness temperature in the source frame is estimated to be $>2.2\times10^{13}$\,K for the blazar core. 
This implies that the inverse Compton limit must be violated in the rest frame of the source, even for the largest Doppler factor $\delta\thicksim25$ reported for 0716$+$714.
\end{abstract}


\section{Introduction} \label{sec:intro}

S5~0716$+$71 (hereafter 0716$+$714) is one of the most active BL~Lac objects and the subject of numerous studies carried out since its discovery in 1979 by \citet{1981AAS...45..367K}. 
It is extremely variable throughout the whole electromagnetic spectrum, with time scales that range from hours to months, including intra-day variability \citep[IDV; e.g.][]{2006AA...451..797O,2014ApJ...783...83L}. 
The duty cycle of the source is about 80\%-90\% \citep[e.g.][]{1995ARAA..33..163W,2014MNRAS.443.2940H}, which implies that it is active almost all the time. It is considered as one of the best candidates for having an intrinsic origin of the observed IDV \citep[e.g.][]{0716Nature,2012MNRAS.425.1357G}, rather than being produced by scintillation in the ionized interstellar medium \citep[ISS,][]{1984AA...134..390R}. 
There are a large amount of evidence supporting this: the observed increase in variability amplitude with frequency \citep{2008AA...490.1019F}; the correlation between optical brightness and radio spectral index, as well as simultaneous change in variability time-scale \citep{1990AA...235L...1W,1996AJ....111.2187W}; the rapid variability at millimeter wavelengths \citep{2006AA...451..797O}; the quasi-periodic intra-hour oscillations at optical band \citep{2010ApJ...719L.153R}; and the highly polarized optical micro flares \citep{2015ApJ...809L..27B}.
However, the observed annual modulation of the variability pattern \citep{2012AA...543A..78L} implies that some of the variability observed in 0716$+$714 is produced indeed by interstellar scintillation, at least at 6~cm and 11~cm \citep[see also][]{2012MNRAS.425.1357G}.

Very long baseline interferometry (VLBI) observations of 0716$+$714 show that the radio continuum emission is dominated by a bright core, with occasional ejection of superluminal components with apparent velocities as high as $25c$ \citep{2005AA...433..815B,2006AA...452...83B, 2009AA...494L...5R, 2017ApJ...846...98J, 2019ApJ...874...43L}.
A joint analysis of the jet kinematic properties and the variability of radio emission in 0716$+$714 yields estimates of Doppler factor $4\leq\delta\leq25$, with bulk jet Lorentz factor $10\leq\Gamma\leq25$ at viewing angle $\theta\leq5^{\degr}$ \citep{2005AA...433..815B,2009AA...494..527H,2017ApJ...846...98J}.

First space VLBI (SVLBI) observations of 0716$+$714, performed with the VLBI Space Observatory Program (VSOP), were reported by \citet{2006AA...452...83B}, revealing that the IDV observed in 0716$+$714 both in the total and polarized emission may originate within the unresolved $<100~\mu$as VLBI core. 
From the variability timescales assuming internal nature of the IDV they estimated a source size of a few $\mu$as, well below the angular resolution provided by VSOP.

\begin{figure}
\includegraphics[angle=270,width=0.41\textwidth]{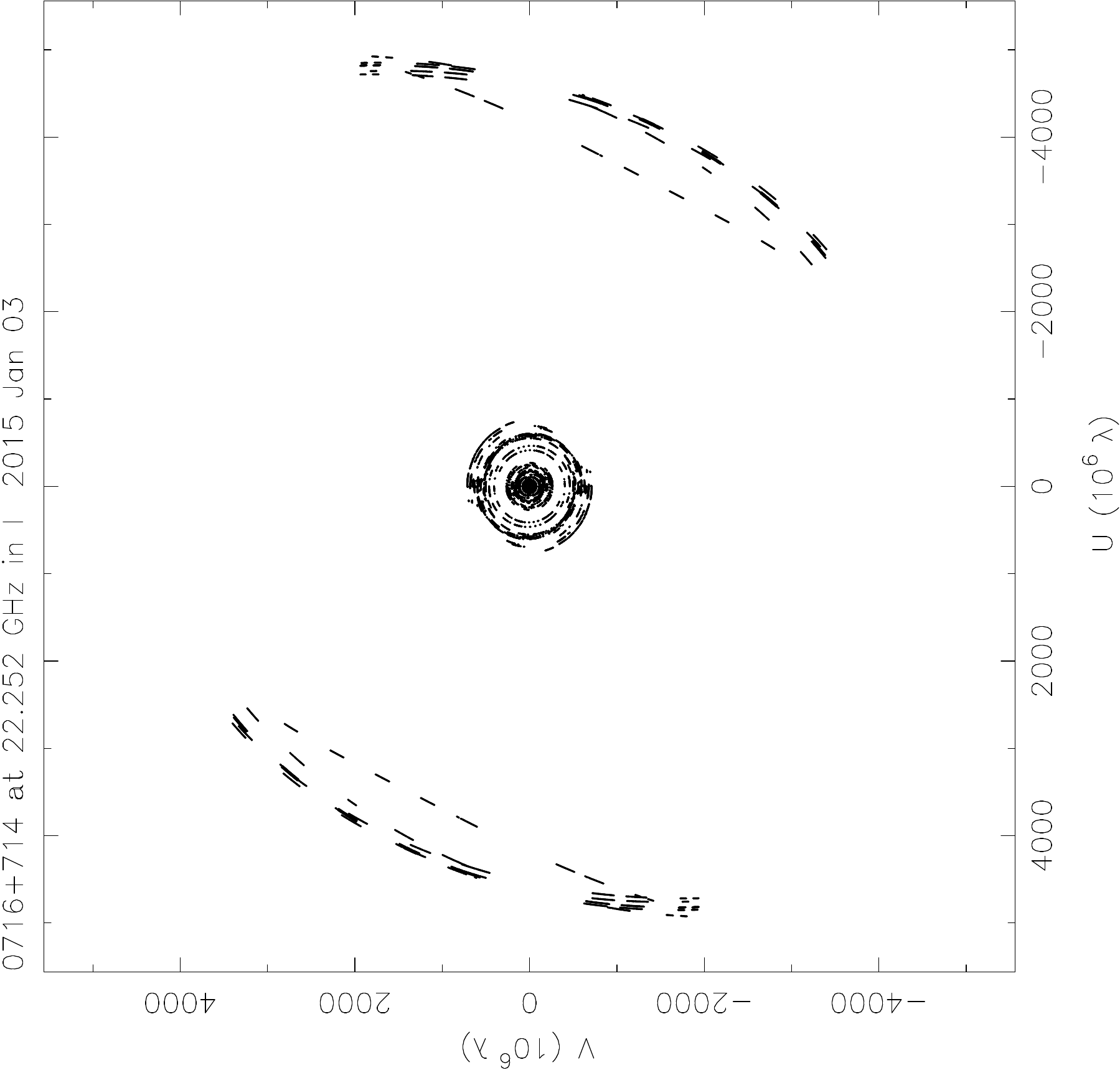}\\
\includegraphics[angle=270,width=0.41\textwidth]{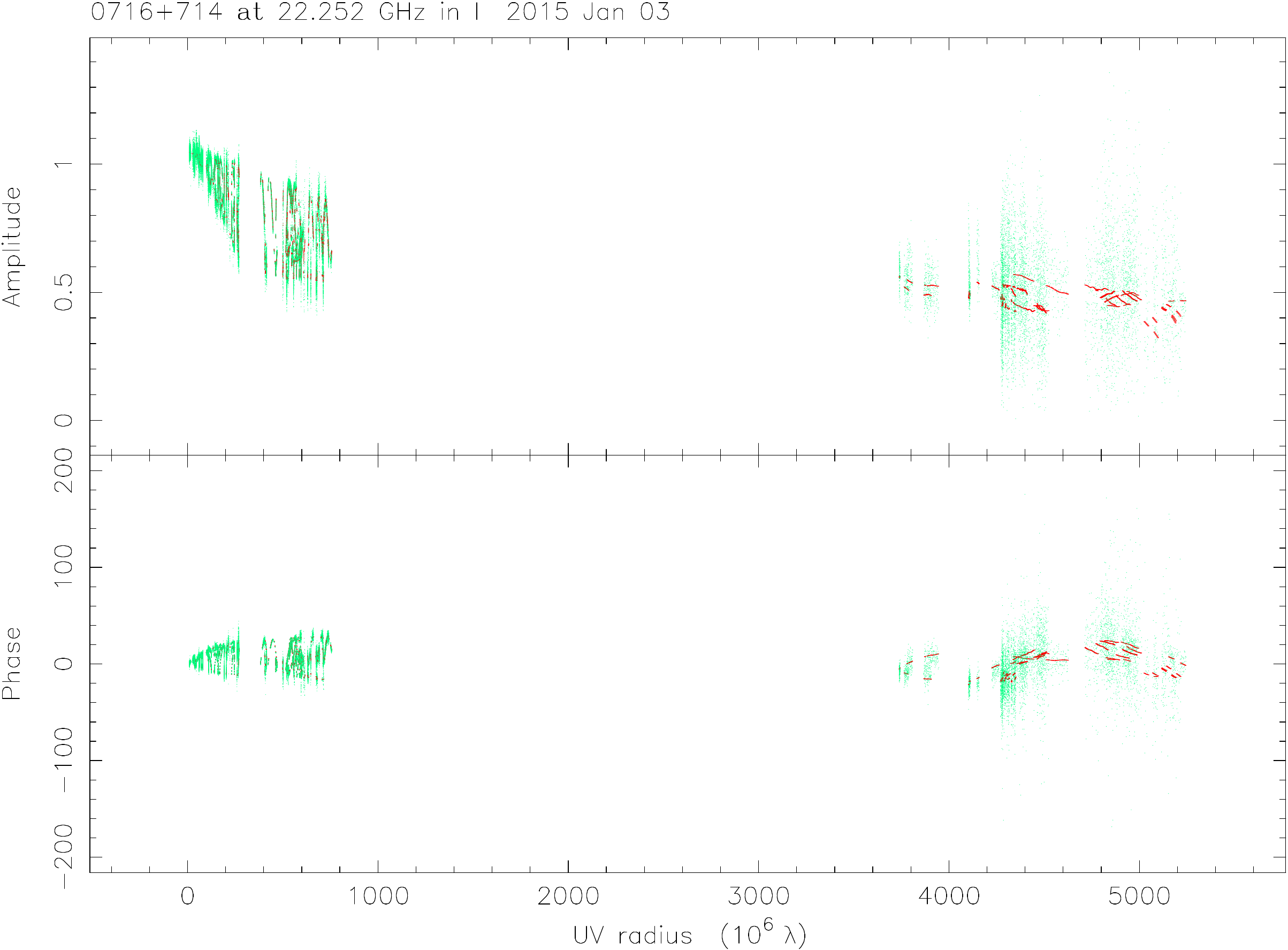} 
\caption{(top) Coverage of the {\em uv}-space Fourier domain for \textit{RadioAstron} detections of 0716$+$714 in 2015 January 3-4 at 22\,GHz. (bottom) Self-calibrated visibility amplitudes (Jy) and phases (degrees) versus {\em uv}-distance for the same experiment. The CLEAN model obtained from hybrid mapping is overlaid in red color. Space-ground baselines start at $\thicksim$3500\,M$\lambda$ ($1\,ED\thicksim940$\,M$\lambda$). 
\label{fig:uvp}}
\end{figure}

Launched in 2011 and observing till January 2019, the \textit{RadioAstron} space VLBI mission \citep{2013ARep...57..153K} featured a 10--m antenna on board the Spektr-R satellite. Revolving around the Earth on an elliptical orbit reaching 360 000 km, it enabled one to probe celestial objects with the maximum angular resolution of $7~\mu$as \citep{2017SoSyR..51..535K}.
The  space  radio  telescope (SRT) operated  in  four  wavelength  bands: P (0.33 GHz, $\lambda = 92$\,cm), L (1.6\,GHz,  $\lambda = 18$\,cm), C (4.8\,GHz,  $\lambda = 6.2$\,cm), and K (18-26\,GHz,  $\lambda = 1.2$--$1.6$\,cm). 
Polarized emission could be studied with \textit{RadioAstron} in the P, L, and K bands in dual circular polarization mode, with the left circularly polarized (LCP) and right circularly polarized (RCP) channels for recording data.

\begin{deluxetable*}{rccccc}
\tablecaption{Radio telescopes participating in {\em RadioAstron} observations of 0716$+$714.} \label{tab:rt}
\tablecolumns{6}
\tablenum{1}
\tablewidth{0pt}
\tablehead{
\colhead{Telescope} & \colhead{Code} & \colhead{$D$} & \colhead{SEFD} & \colhead{BW}& \colhead{$B_\mathrm{max}$} \\
\colhead{} & \colhead{} & \colhead{(m)} & \colhead{(Jy)} & \colhead{(MHz)}& \colhead{(ED)}}
\startdata
Spektr-R (Space)    & RA & 10  & 4670$^{a}$, 3680$^{b}$ & 32 &4.96\\
Brewster (USA)      & BR & 25  & 640  & 64  & 4.81\\
Effelsberg (Germany)& EF & 100 & 90   & 128 & 4.68\\
Ford Davis (USA)    & FD & 25  & 640  & 64  & 4.93\\
Green Bank (USA)    & GB & 100$^{c}$  & 30  & 64 & 4.96\\
Hancock (USA)       & HN & 25  & 640  & 64  & 0.40\\
Los Alamos (USA)    & LA & 25  & 640  & 64  &4.91\\
Noto (Italy)        & NT & 32  & 800  & 128 &  0.66\\
Owens Valley (USA)  & OV & 25  & 640  & 64  & 4.81\\
Pie Town (USA)      & PT & 25  & 640  & 64  & 4.91\\
Shangai (China)     & SH & 65  & 1700 & 128 & 0.72\\
Torun (Poland)      & TR & 32  & 500  & 128 & 0.59\\
\enddata 
\tablecomments{$D$: antenna diameter ($^{c}$: equivalent diameter); SEFD: system equivavelnt flux density at 22\,GHz ($^{a}$ for LCP, $^{b}$ for RCP); BW: observing bandwidth per each circular polarization; $B_\mathrm{max}$: largest projected baseline in detections, in units of Earth diameter.}
\end{deluxetable*}

Blazar 0716$+$714 was the first AGN imaged with \textit{RadioAstron}. Observations were made on March 14--15 2012 at 4.8\,GHz during the Early Science Program period \citep{2013ARep...57..153K}. One of the \textit{RadioAstron} Key Science Program (KSP) continued these experiments, particularly focusing on the polarimetry of the most active and highly polarized AGNs in the sky. 
The Polarization KSP has collected data throughout the whole {\em RadioAstron} science program.
The first test polarimetric observations were performed on 2013 March 9-10 at 18~cm targeting the quasar TXS~0642$+$449 \citep{2015AA...583A.100L}. 
The total and linearly polarized emission was detected at baselines up to 75 560 km ($\sim6$~Earth diameters, ED), resulting in angular resolution of 0.8~mas. 
It was shown, that instrumental polarization (leakage between the two polarization channels) of the space antenna is smaller than 9\%, demonstrating the \textit{RadioAstron} polarization capabilities for high fidelity polarization imaging \citep[see also][]{2015CosRe..53..199P}.
The first polarimetric experiment at 1.3~cm took place on 2013 November 10 \citep{2016ApJ...817...96G}, when BL~Lacertae was detected on the projected baselines up to $\sim8$~ED, delivering the image with the highest achieved to date angular resolution of $21~\mu$as.
The instrumental polarization of the SRT was found to be less than 9\%, 
demonstrating the possibility of carrying out polarization measurements with {\em RadioAstron} at its highest observing frequency.
Combination of \textit{RadioAstron} 22\,GHz and quasi simultaneous ground-based 15 and 43\,GHz images resulted in detecting a gradient in rotation measure and intrinsic polarization vector as a function of position angle with respect to the core of BL~Lacertae, suggesting that the blazar jet is threaded by helical magnetic field.
The intrinsic de-boosted brightness temperature in the unresolved core of the source was estimated to be $>3\times10^{12}$~K, in excess of theoretical limits \citep{1969ApJ...155L..71K, readhead_94}.

Recently, \citet{2017AA...604A.111B} presented the results of the SVLBI imaging experiment of the quasar 3C273 on 2014 November 18-19 at 22\,GHz. 
The observations probed the source during a particularly low activity state, yielding brightness temperature at the level of $5\times10^{9}$~K. 
\citet{2016ApJ...820L...9K} report almost two orders of magnitude higher value of $1.4\times10^{13}$~K, based on the \textit{RadioAstron} observations of the quasar on 2013 February~2 at the same frequency.
\citet{2017AA...604A.111B} conclude that detected extreme brightness temperature in the quasar seen by \citet{2016ApJ...820L...9K} one year before their study represents a short-lived phenomenon caused by a temporary departure from the equipartition between the radiating particles and magnetic field energy density. 
This implies that the source state plays a principal role in the characterization of the observable physical quantities.

This paper continues a series of studies of AGN jets within the \textit{RadioAstron} Polarization KSP: we present here our observations of 0716$+$714 probing the jet structure at the finest angular resolution possible to investigate the origin of the IDV and reconstruct the magnetic field in the vicinity of the central black hole.

The 0716$+$714 has no identified spectroscopic redshift because of featureless optical continuum and bright nucleus in the optical.
Most recent attempts to measure the blazar redshift have constraint it within a range of $0.2315<z<0.3407$ \citep{2013ApJ...764...57D}, based on the detection of Ly-$\alpha$ systems in the far ultra-violet spectrum of the source. 
This is consistent with the photometric detection of the blazar's host galaxy at $z=0.31\pm0.08$ \citep{2008A&A...487L..29N}. 
Recently \citet{2018AA...619A..45M} used simultaneous spectra from MAGIC and \textit{Fermi}-LAT at energies above 0.1~GeV and estimated the blazar redshift to be $z =0.31\pm0.05$. 
We adopt this value throughout the paper.

We assume the flat $\Lambda$CDM cosmology with the matter density $\Omega_\mathrm{m}=0.3$, cosmological constant $\Omega_{\Lambda}=0.7$ and Hubble constant $H_0 = 70$km s$^{-1}$Mpc$^{-1}$, \citep{2014ApJ...794..135B,2016AA...594A..13P}, which corresponds to the luminosity distance $D_\mathrm{L}=1.6$~Gpc and gives a scale of 4.56~pc per 1~mas at the redshift of 0.31. 

\section{Observations and data reduction} 
\label{sec:obs}

\subsection{\textit{RadioAstron} Imaging experiment}
\label{sec:observ_ra}

The imaging experiment on 0716$+$714 was performed in 2015 January 3-4 at 22.2\,GHz during a twelve-hour space VLBI session (\textit{RadioAstron} project code raks11aa, global VLBI project code GL041A).
Out of 20 scheduled ground stations, 8 were lost due to different technical problems: Mets\"ahovi (Finland), Robledo (Spain), Jodrell Bank (UK), Sardinia (Italy), Yebes (Spain), North Liberty (USA), Kitt Peak (USA), St.~Croix (USA), and Mauna Kea (USA).
Twelve ground antennas delivered data which could be used for the correlation. 
They are listed in Table~\ref{tab:rt} together with their basic parameters.

The data were recorded in two circular polarizations (right and left, RCP and LCP) starting from the frequency of 22220\,MHz in a bandwidth of 32, 64 and 128\,MHz per polarization, depending on the radio antenna (See Table~\ref{tab:rt} for details).
Each bandwidth was split into two, four and sixteen 16\,MHz intermediate frequency (IF) channels, respectively.
Two calibrators, 4C$+$39.25 and 0836$+$710, were observed only by the ground array during the gaps required to cool down the motor drive of the on-board high-gain antenna of the Spektr-R satellite. 
These gaps were also used to schedule several scans at 15 and 43 GHz with the VLBA stations in order to provide a quasi-simultaneous multi-frequency coverage of our target. 
Unfortunately, due to the loss of the data at four of the VLBA antennas, the resulting 15 and 43\,GHz datasets were insufficient to obtain reliable images at these frequencies.

The VLBI experiment was correlated using the upgraded \textit{ra} version of the DiFX correlator developed at the Max-Planck-Instit\"ut f\"ur Radioastronomie (MPIfR) in Bonn \citep{2016Galax...4...55B}. 
At the correlation stage, a rough fringe search on ground-to-space baselines was performed starting from the scans nearest to the spacecraft perigee, corresponding to a projected baseline of about 4~ED for this experiment. 
A significant interferometric signal was found up to about 5~ED. For the remaining scans solutions for delay and delay-rate were extrapolated from the previous ones in order to optimize the centering of the correlation window.

Data reduction and hybrid imaging were performed in AIPS \citep{aips} and Difmap \citep{1994BAAS...26..987S, 1997ASPC..125...77S}. 
Following the procedure for calibrating \textit{RadioAstron} data described in \cite{2016ApJ...817...96G}, we first performed a fring-fitting of the ground array, and then coherently phased it to search for the fring solutions of the space antenna. 
We were able to find significant fringes between ground antennas and the space telescope at the maximum projected baseline of 5.25\,G$\lambda$, or 5.56\,ED (70\,833\,km). 
No fringes to the SRT were found for two scans at about 6~ED despite integrating the signal from the two polarizations and IFs.
We considered the a~priori visibility amplitude accuracy at the level of 13\%.
The delay difference between the two polarization was corrected using the task RLDLY in AIPS.
The instrumental polarization (or leakage) was derived with the task LPCAL in AIPS, using our target 0716$+$714 since it is the only source observed with the Spektr-R radio antenna; besides, it contains the best parallactic-angle coverage among all the observed sources.
Calibration of the absolute EVPA orientation was obtained through comparison with the quasi-simultaneous observations of our target and calibrators obtained with Effelsberg single dish at 2.8~cm and 6~cm (epoch 2015--01--01), VLBA-MOJAVE at 2~cm (2015--01--18) and VLBA-BU at 7~mm (2014--12--29 and 2015--02--14).
We estimate our absolute EVPA values to be accurate within 6\degr.

\begin{figure*}
\includegraphics[angle=270,width=0.31\textwidth]{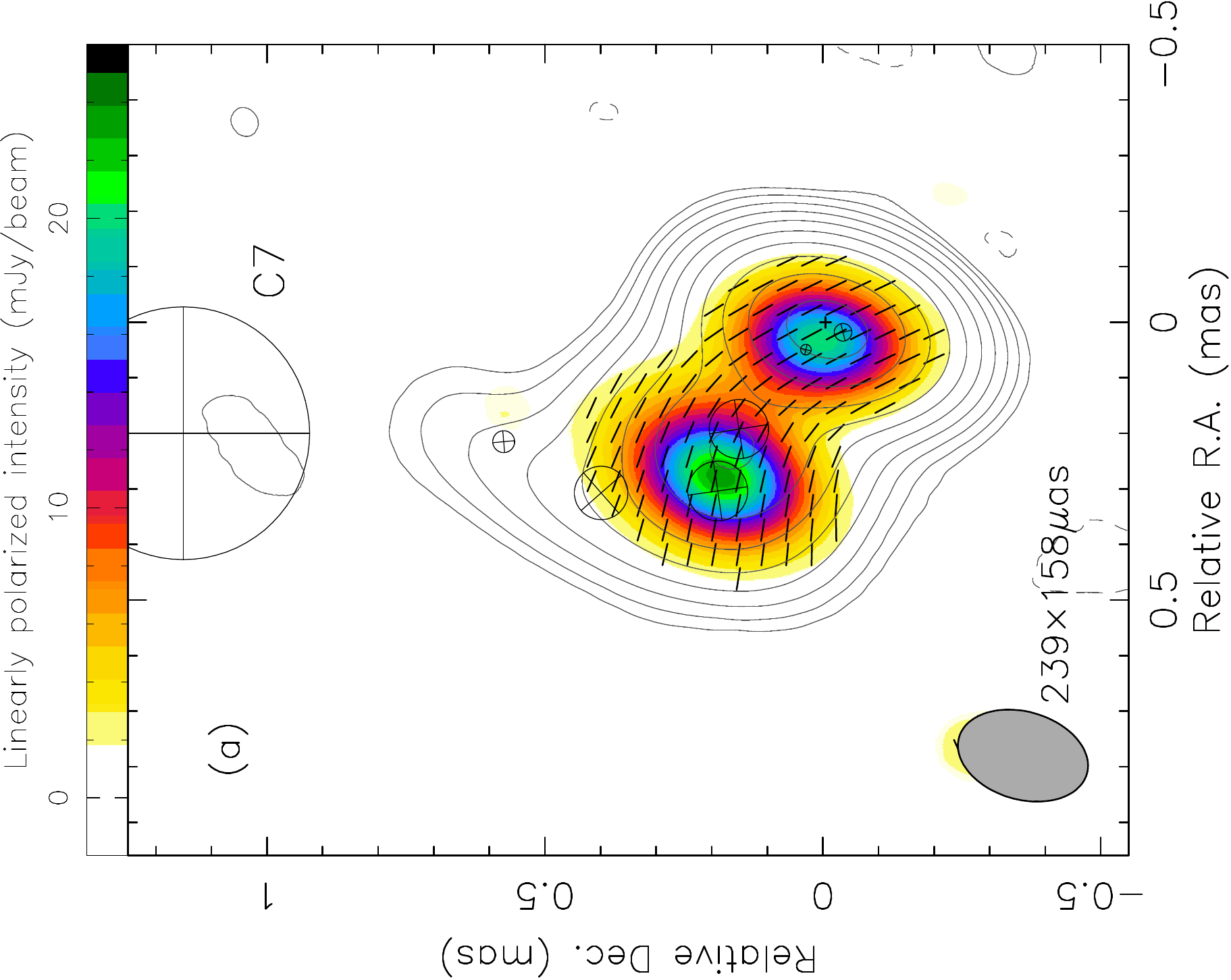} 
\includegraphics[angle=270,width=0.31\textwidth]{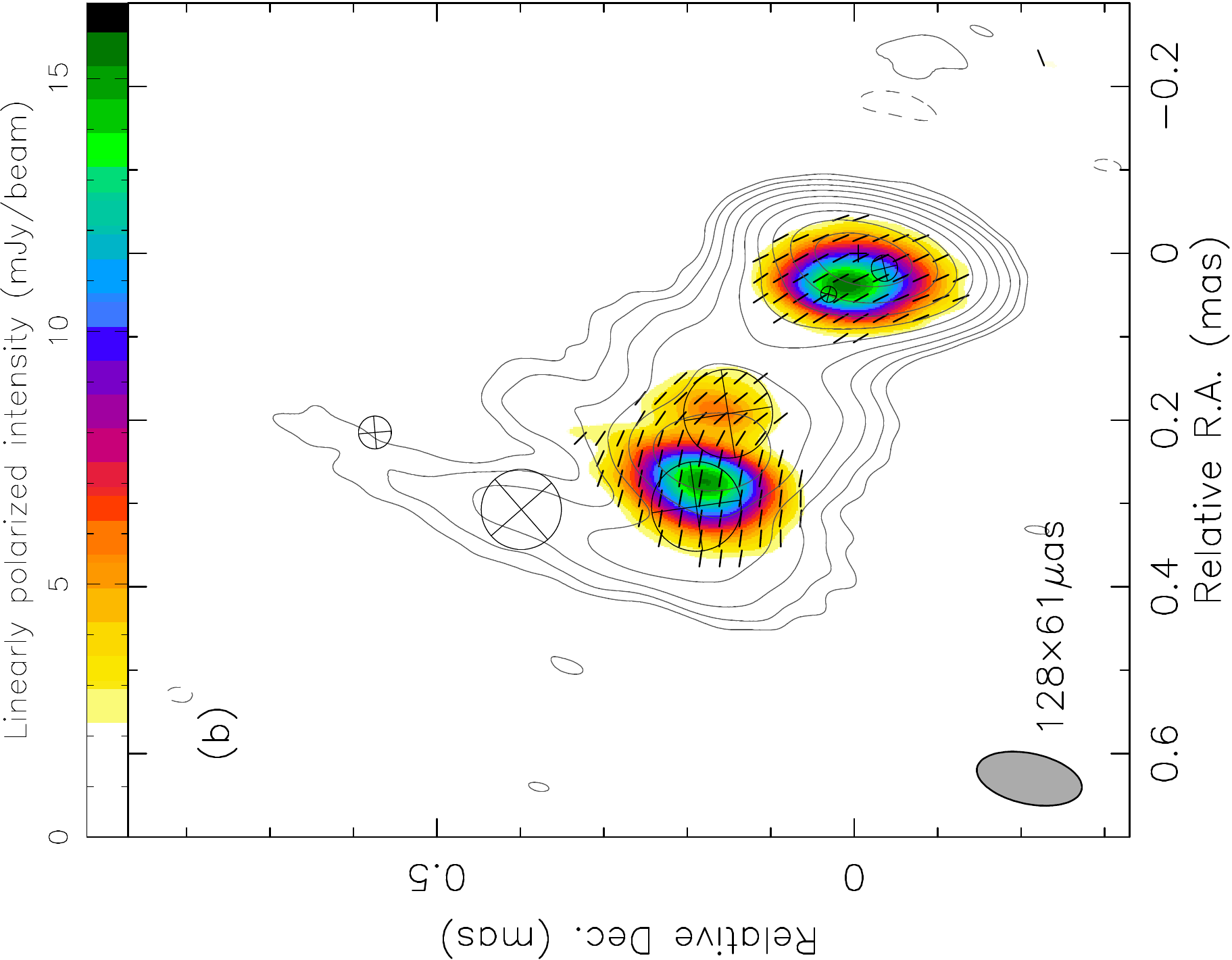} 
\includegraphics[angle=270,width=0.31\textwidth]{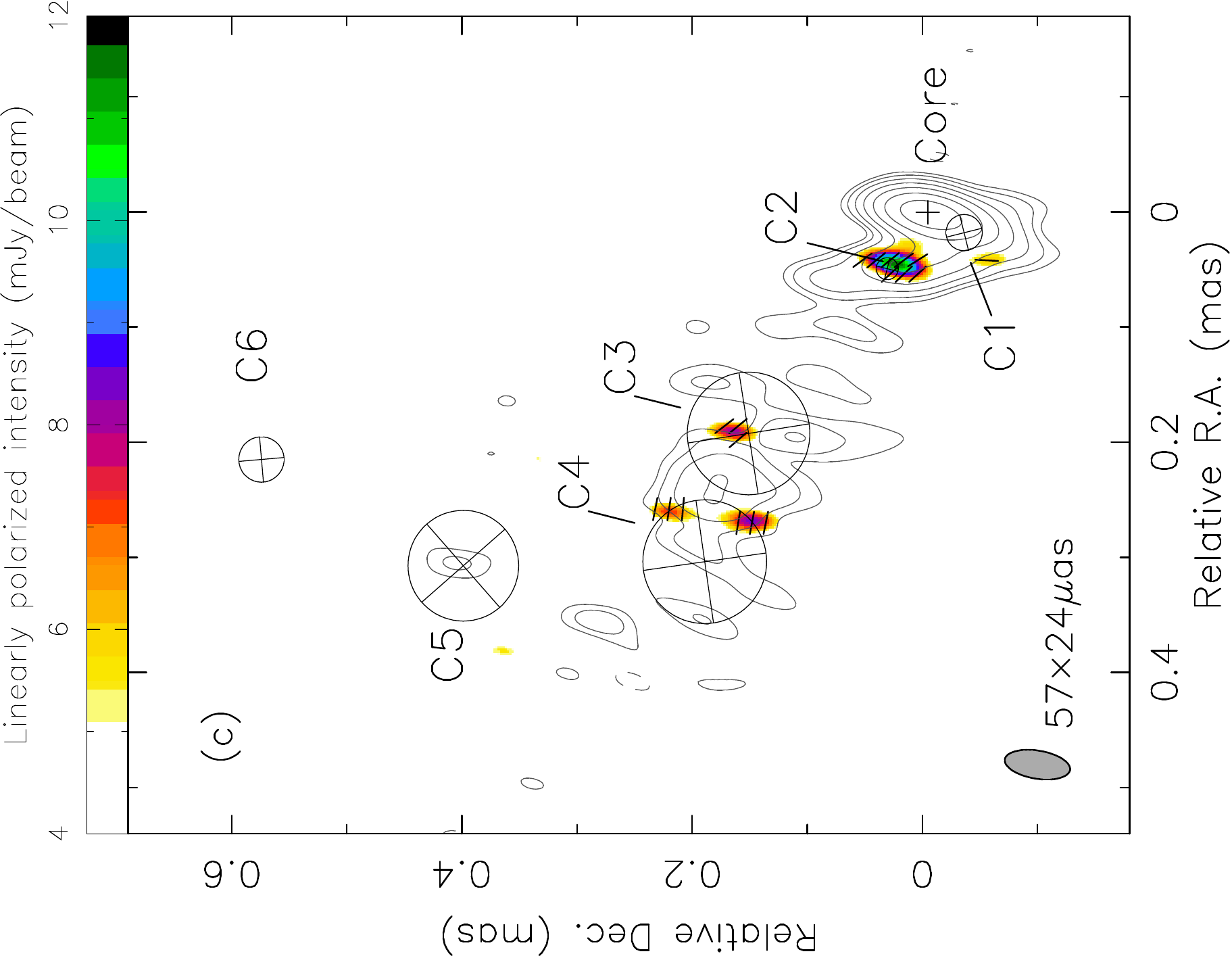}\\ 
\includegraphics[angle=270,width=0.31\textwidth]{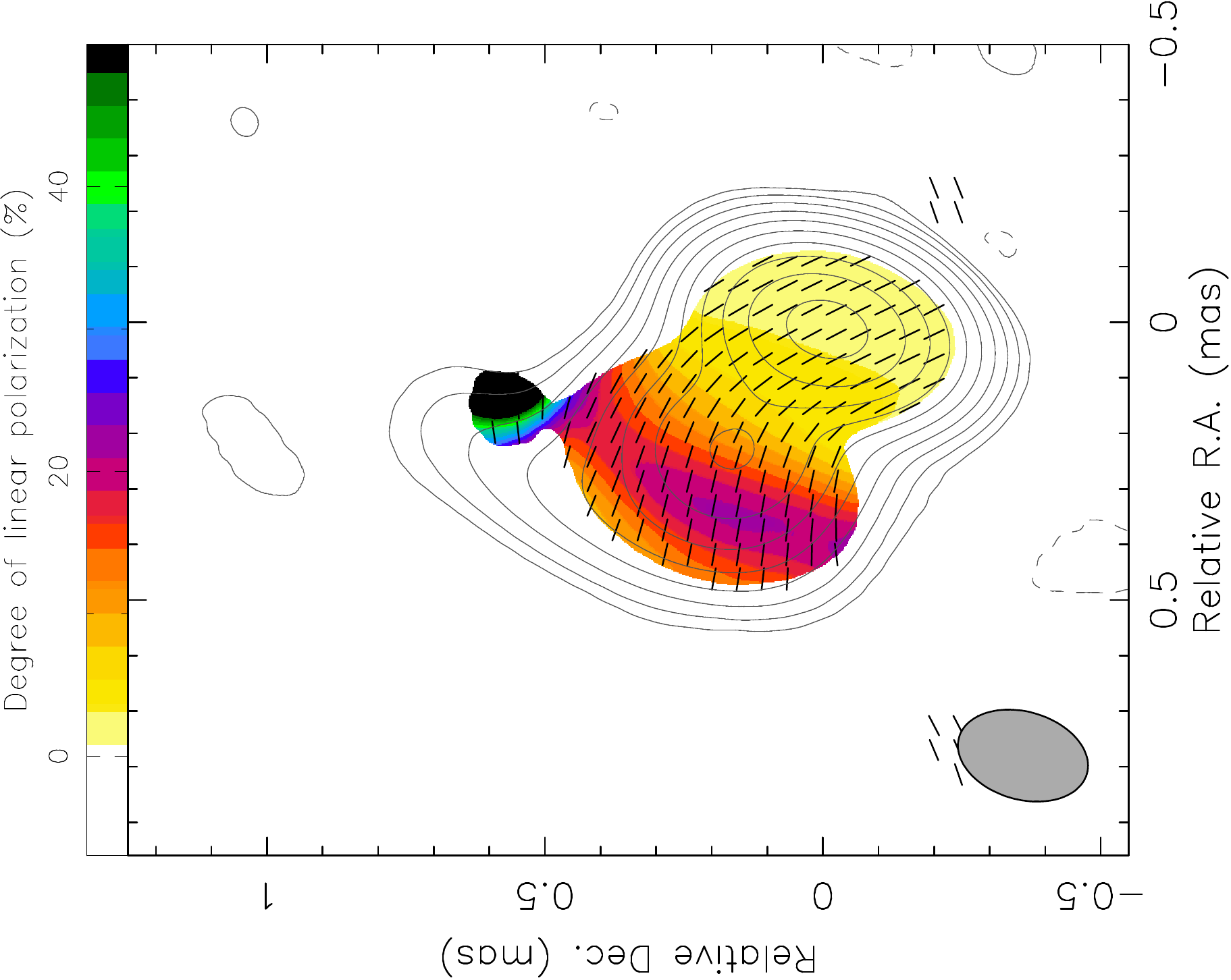} 
\includegraphics[angle=270,width=0.31\textwidth]{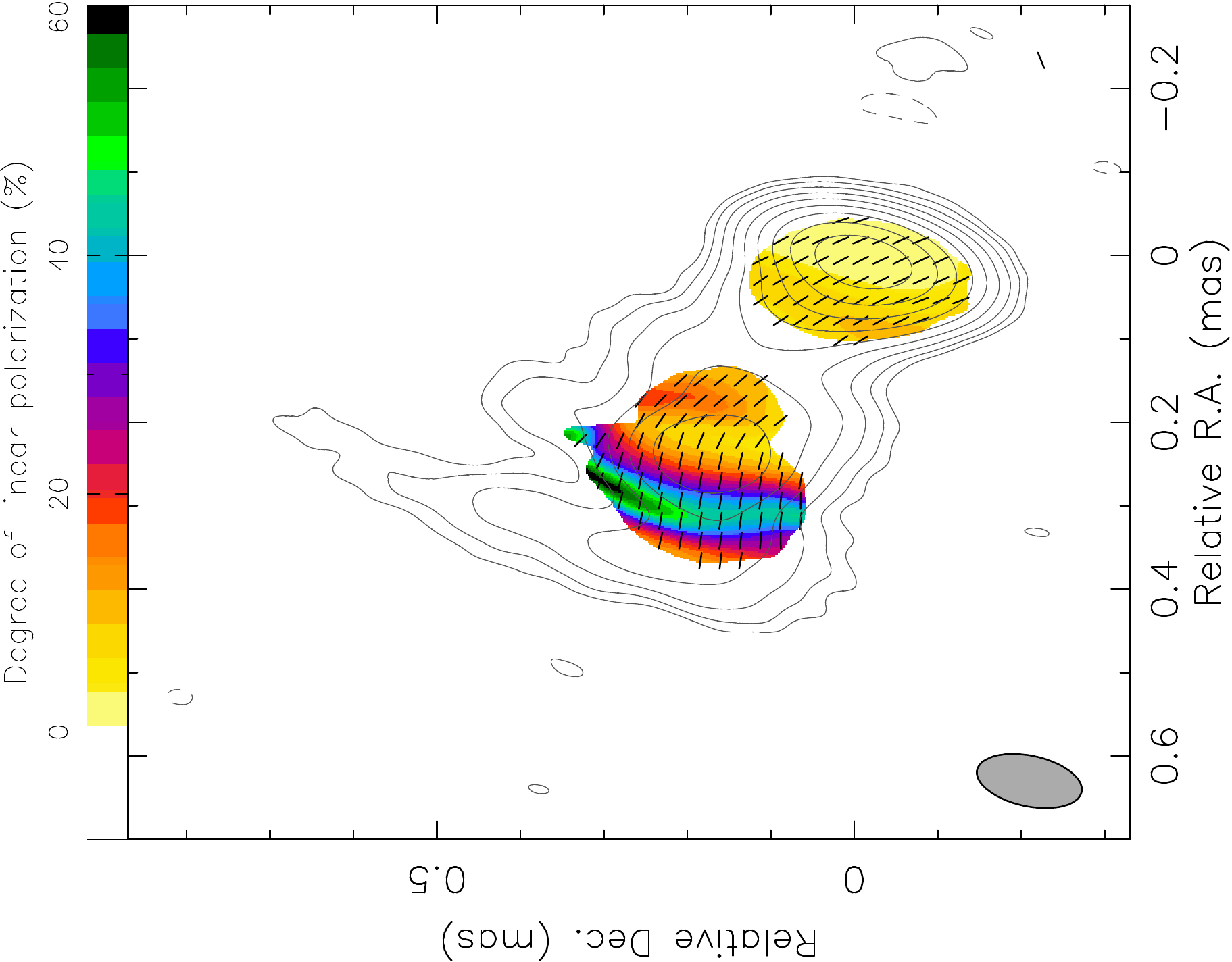} 
\includegraphics[angle=270,width=0.31\textwidth]{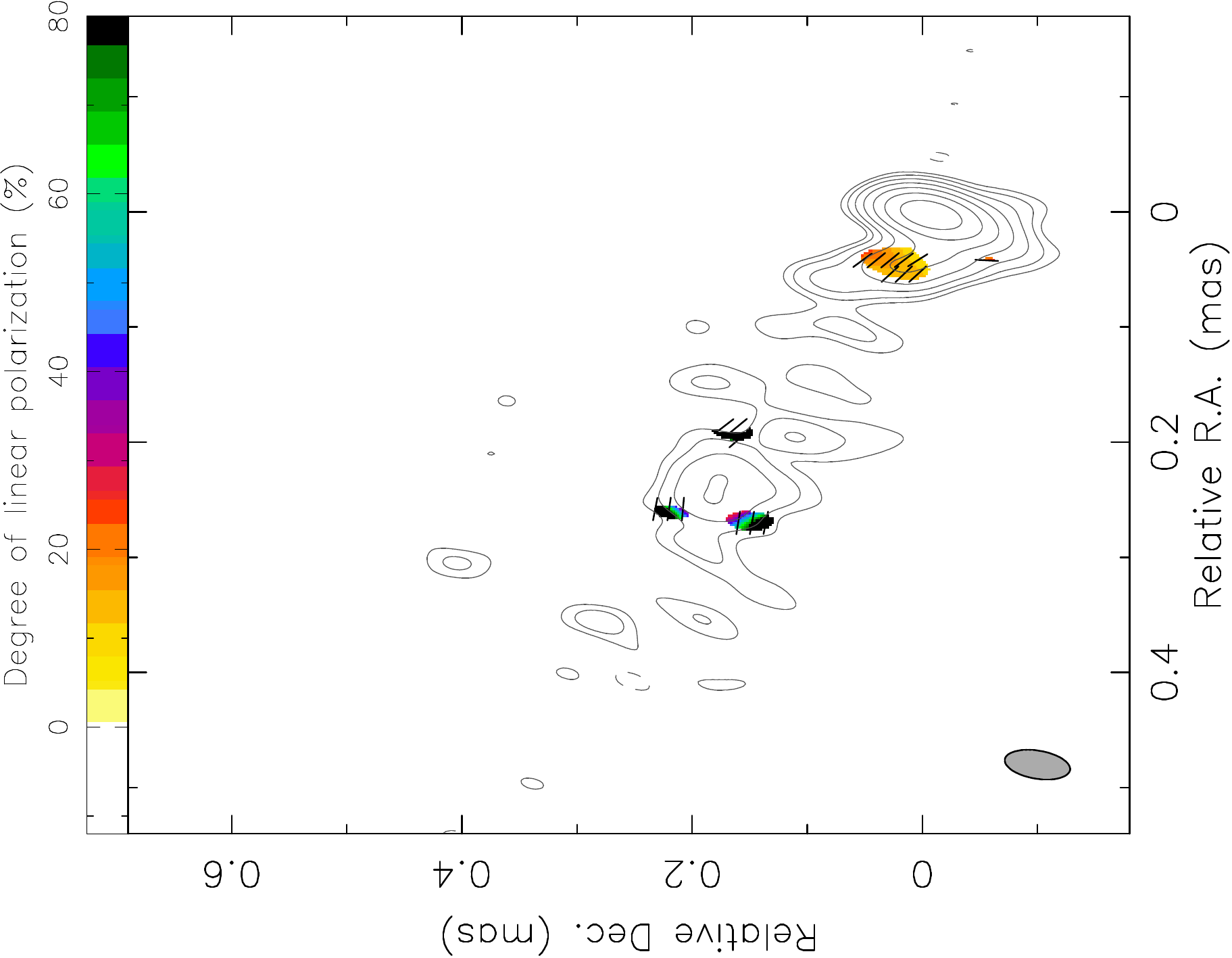}\\
\caption{(upper row) \textit{RadioAstron} polarimetric and total intensity SVLBI image of 0716$+$714 obtained on 2015 January 3-4 at 22\,GHz with (a) natural, (b) uniform and (c) super-uniform weighting. The total intensity contours are the following (a) $-$0.8, 0.8, 1.7, 3.4, 6.8, 13.5, 27.0, 54.1, 108.1, 216.3, 432.6~mJy $\mathrm{beam}^{-1}$; (b) $-$1.4, 1.4, 2.8, 5.6, 11.2, 22.3, 44.6, 89.2, 178.5, 356.9~mJy $\mathrm{beam}^{-1}$ and (c) $-$4.5, 4.5, 8.9, 17.8, 35.7, 71.3, 142.6, 285.3 $\mathrm{beam}^{-1}$ Linearly polarized intensity is shown by color, the value of $\sigma_\mathrm{rms}$ corresponds to (a) 0.26, (b) 0.30 and (c) 1.51 ~mJy $\mathrm{beam}^{-1}$.  Synthesized beam is given at the position angle of (a)$-15\fdg9$, (b) $-11\fdg8$ and (c) $-9\fdg3$, and is displayed by shaded ellipse at the left bottom corner. Circles and crosses indicate the positions and the sizes (FWHM) of Gaussian components, which were model fitted to the visibility data in the {\em uv}-plane. (bottom row) Same as above , but for the degree of linear polarization.
\label{fig:imspace}}
\end{figure*}

\subsection{Single dish monitoring at 37 GHz}

Mets\"{a}hovi Radio Observatory has a dedicated program to monitor a sample of extragalctic radio sources, inlcuding bright AGNs, that has been running since 1980 using 13.7-m diameter antenna \citep{1998AAS..132..305T}. 
The program comprises regular observations of the total flux density mainly at 22 and 37\,GHz.
We made use of these 37\,GHz data in our analysis.

\subsection{43 GHz VLBA observations}

Our study makes use of 43\,GHz VLBA data from the VLBA-BU-BLAZAR monitoring program\footnote{\url{http://www.bu.edu/blazars/VLBAproject.html}}. 
A total of 50 epochs of 0716$+$714 we used, covering the period from 2012 January 27 until 2017 April 16, with a monthly cadence approximately. 
Further details about the VLBA-BU-BLAZAR program and the calibration of these data can be found in \citet{2017ApJ...846...98J}.

\section{Results} 
\label{sec:res}

\subsection{22 GHz SVLBI image and jet structure}
\label{s:svlbi}

Figure~\ref{fig:uvp} shows the coverage in the visibility domain, the so-called {\it uv}-coverage, of the fringe-fitted solutions for our \textit{RadioAstron} observations of 0716$+$714 in 2015 January 3-4 at 22~GHz. 
The resulting polarization space VLBI images of the blazar are shown in Fig.~\ref{fig:imspace} using three different weightings: natural, uniform, and ``super'' uniform, when the gridding weights of the visibilities for the longest baseline and particularly for the {\em RadioAstron} are not scaled by visibility amplitude errors and thus increase from natural to super uniform weightings.
This therefore yields the higher angular resolution, achieving of about $24~\mu$as, but lower image sensitivity for the super uniformly weighted image.
Nonetheless, \citet{2012A&A...541A.135M} show that the over-resolution power of an interferometer (the minimum possible size of a source whose structure can be detected) is a function of its sensitivity, implying that an interferometer is capable of resolving structures below the diffraction limit $\lambda/B_\mathrm{max}$ (known as superresolution).
Considering the emission in the inner 100\,$\mu$as of the 0716$+$714 jet, we estimate the signal-to-noise ratio $SNR$ at the level of $\thicksim30-90$. 
Following \citet{2012A&A...541A.135M} and considering the uniformly weighted beam size $\theta_\mathrm{beam}$, it enables us to probe the details of the jet down to $\thicksim\theta_\mathrm{beam}/SNR^{0.5}\thicksim$ few $\mu$as, justifying usage of the superresolution.

\begin{figure}
\includegraphics[angle=0,width=0.46\textwidth]{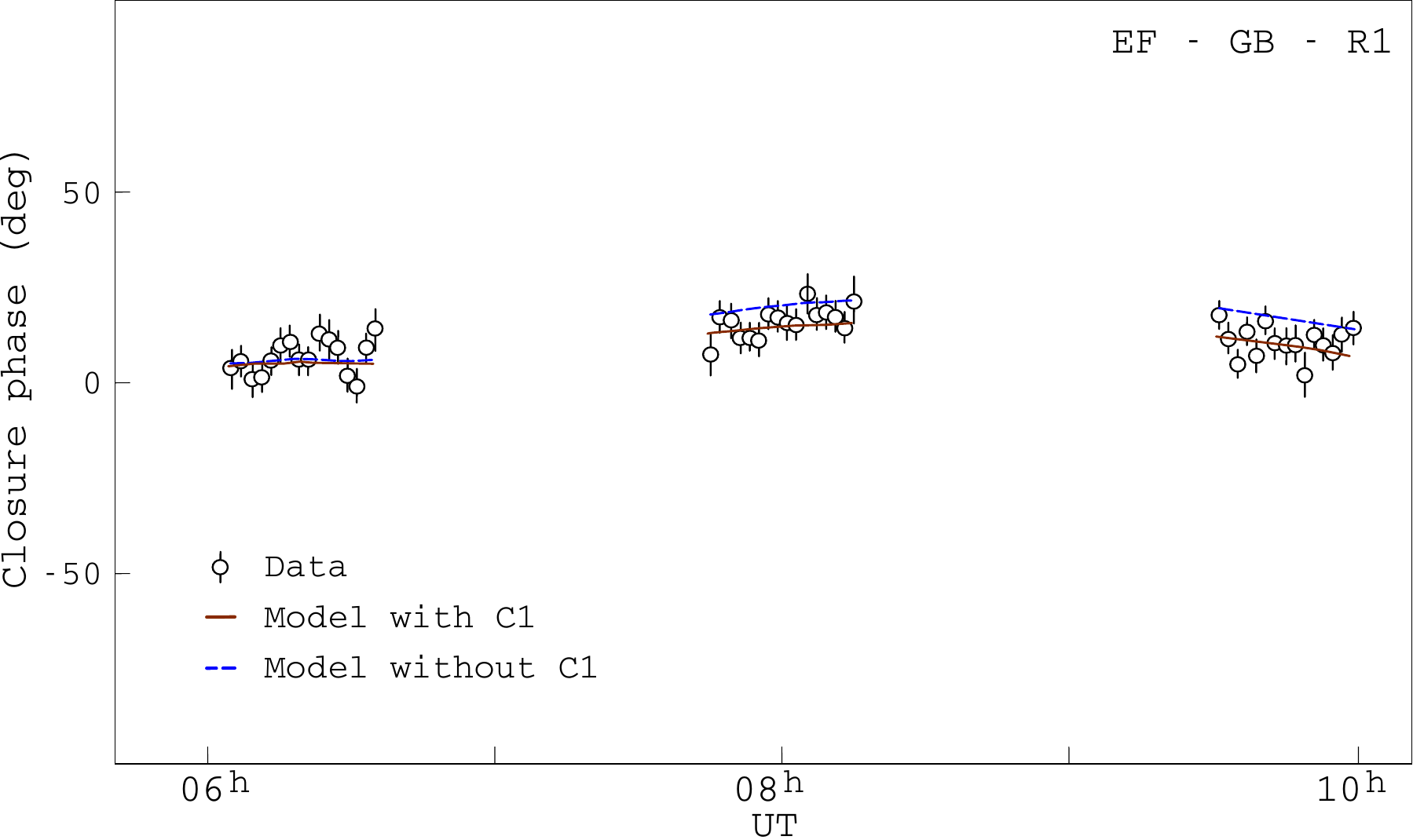}
\caption{Selected, non-zero closure phases of 0716$+$714 as a function of time formed on the triangle {\em RadioAstron}--Effelsberg--Green Bank, along with the fitted models with and without component C1 discussed in Sect.~\ref{s:svlbi}. $\chi^2$ value evaluating the model fit with component C1 is 2.8 times lower than for the model without component C1.
\label{fig:cp}}
\end{figure}

To build a {\em RadioAstron} image, we made use of closure quantities, as they are helpful in constraining the overall source structure \citep{1974ApJ...193..293R, 1988AJ.....95.1278R}. Particularly, closure phase represents a directed sum of visibility phases around an interferometric triangle, which cancels out the station-based phase errors on each individual visibility \citep{1958MNRAS.118..276J}.
We computed the $\chi^2$-statistics for the closure phase, $\chi^2_\mathrm{CP}$, and closure amplitude, $\chi^2_\mathrm{log CA}$, to quantify the agreement between the reconstructed image and the data following \citet{2019ApJ...875L...4E}.
We have compared these quantities for the models with and without the extension of the core in the southeast direction.
It turned out that the model without the extension has the $\chi^2$ values higher by a coefficient of 2.8 and 2.4 than the model with the extension for the closure phases and amplitudes, respectively.
Among a few alternatives the presented 0716$+$714 image provides the best overall fit to the data, and especially to the non-zero closure phases between {\it RadioAstron}, and the most sensitive ground radio antennas, namely Green Bank and Effelsberg (See Fig.~\ref{fig:cp}).
We examined an effect of amplitude calibration of the space antenna on these results.
Variation of the amplitude correction within 10\% yields no significant changes, supporting robustness of the obtained 0716$+$714 image and extension of the core in particular.
Nevertheless, future space and mm-VLBI observations at similarly high angular resolution as those presented here, now possible thanks to the Event Horizon Telescope \citep{2019ApJ...875L...2E}, would be required to test the reliability of the obtained 0716$+$714 model.

\begin{deluxetable*}{lllllll}
\tablecaption{Results of Gaussian model fitting of the 22\,GHz SVLBI data \label{tab:mdf}}
\tablecolumns{6}
\tablenum{2}
\tablewidth{0pt}
\tablehead{
\colhead{Comp.} & \colhead{Flux density} & \colhead{Distance} & \colhead{PA} & \colhead{Size}& \colhead{T$_\mathrm{b,rf}$} \\
\colhead{} & \colhead{(mJy)} & \colhead{($\mu$as)} & \colhead{(\degr)} & \colhead{($\mu$as)}& \colhead{(K)}
}
\startdata
Core& 429$\pm$56 & 4.8$\pm$0.2 & 176$\pm$6    & $<12\times$5& $>2.2\times10^{13}$\\
C1 & 212$\pm$28  & 40.5$\pm$0.5& 153.7$\pm$0.7& 31.5$\pm$0.4   & $(7.1\pm0.9)\times10^{11}$\\
C2 & 142$\pm$19  & 58.0$\pm$0.5& 58.3$\pm$0.3 & 19.1$\pm$0.4   & $(1.29\pm0.18)\times10^{12}$\\
C3 & 146$\pm$19  & 245$\pm$3   & 51.9$\pm$0.3 & 106.6$\pm$0.8  & $(4.3\pm0.6)\times10^{10}$\\
C4 & 94$\pm$13   & 358$\pm$3   & 58.1$\pm$0.2 & 107.5$\pm$1.6  & $(2.70\pm0.4)\times10^{10}$\\
C5 & 12.0$\pm$1.6& 503$\pm$6   & 37.6$\pm$0.5 & 96$\pm$7       & $(4.3\pm0.8)\times10^{9}$\\
C6 & 5.6$\pm$0.8 & 613$\pm$6   & 20.5$\pm$0.5 & 39$\pm$12      & $(1.1\pm0.7)\times10^{9}$\\
C7 & 5.8$\pm$0.8 & 1168$\pm$11 &  9.8$\pm$1.3 & 455$\pm$12     & $(9.2\pm1.3)\times10^{7}$\\
C8 & 3.8$\pm$0.5 & 1878$\pm$5  & 22.1$\pm$0.2 & 211$\pm$4      & $(2.8\pm0.4)\times10^{8}$\\
\enddata
\tablecomments{The column identifiers correspond to: ``Comp.'' - component label, ``Flux'' - flux density, ``Distance'' - distance from the map centre, ``PA'' - position angle of the component from the map centre, ``Size'' - component FWHM, ``T$_\mathrm{b,rf}$'' - rest frame brightness temperature (i.e. corrected for redshift and not corrected for Doppler boosting, see eq.~\ref{eq:tbgr}).}
\end{deluxetable*}

The observed brightness distribution in the jet was represented by a number of circular two-dimensional Gaussian components, which were fitted to the visibility data, using task \textit{modelfit} in \texttt{Difmap}. 
These model-fitted Gaussian components are shown in Fig.~\ref{fig:imspace} and their parameters are listed in Table~\ref{tab:mdf}. The standard deviations for the fitted flux density, position, and size were calculated using the {\em showmodel} routine in \texttt{Difmap}.

The {\em RadioAstron} image in the right panel of Fig.~\ref{fig:imspace} shows a complex, bent structure in the central 0.1~mas core region consisting of an unresolved core and nearby components C1 and C2, located at 40.5~$\mu$as and 58.0~$\mu$as from the core, respectively. 
The jet initially extends towards the south-east, where component C1 is located at a position angle of 153.7\degr, followed by a sharp bending of about 95\degr\, towards the north-east, maintaining that direction for about 1~mas until another sharp bend towards the north-west is observed.
For the unresolved VLBI core we have estimated an upper limit of its size following \citet{2005astro.ph..3225L}, yielding $\theta_\mathrm{core} < 12\times 5~\mu$as.

The VLBA image of 0716$+$714 made at 1.4\,GHz by the MOJAVE program\footnote{Partial release of the 18--22\,cm MOJAVE observations \url{http://www.physics.ucc.ie/radiogroup/18-22cm_observations.html}. The image is available at \url{http://www.physics.purdue.edu/astro/MOJAVE/sourcepages/0716+714.shtml}} shows that jet changes its PA from about $26^\circ$ to about $-50^\circ$ at a distance of about 70~mas from the core. 
Progressively larger-scale VLA image of \citet{2000MNRAS.313..627G} obtained at 5\,GHz shows that the 0716$+$714 jet extends more than 4\,arcsec downstream, at the jet PA of $\thicksim-60^{\circ}$.
Considering the direction of both the component C1 and the arcsec-scale jet relative to the core, it yields about 210\degr\, offset between their orientations in projection on the sky plane. 

As 0716$+$714 is oriented close to the line of sight, intrinsic variations of the jet PA should be significantly amplified in projection of the sky. Assuming the jet viewing angle of $\theta\leq5^{\circ}$, the apparent bend then corresponds to an actual change of $210^\circ\times\mathrm{sin}\theta\lesssim10^{\circ}$. See e.g. \citet{2006ApJ...647..172S} for discussion of possible mechanisms capable of producing curved structures in AGN jets.

Stacked-epoch analysis of VLBA images shows that the intrinsic openning angle of 0716$+$714 jet is $1.6\degr\pm0.2\degr$ \citep{2017MNRAS.468.4992P}. Thus, we can observe the inner jet of the source at viewing angle smaller than its opening angle, i.e. directly inside the outflow \citep{2013AJ....146..120L}. Therefore, inner jet may appear bent as individual emerging features are ejected at different PAs \citep[e.g.][]{2015AA...578A.123R} and fill the entire width of the jet.

\subsection{Linear polarization and rotation measure}

The naturally weighted and uniformly weighted \textit{RadioAstron} images in Fig.~\ref{fig:imspace} show that the linearly polarized emission of 0716$+$714 is dominated by two main features: one associated with the core area and a second one located in the jet, near component C4. 
The super-uniformly weighted image in the right panel of Fig.~\ref{fig:imspace} shows weakly polarized structures located near the component C2 and farther downstream in the jet.

\begin{figure*}
\includegraphics[angle=270,width=0.29\textwidth]{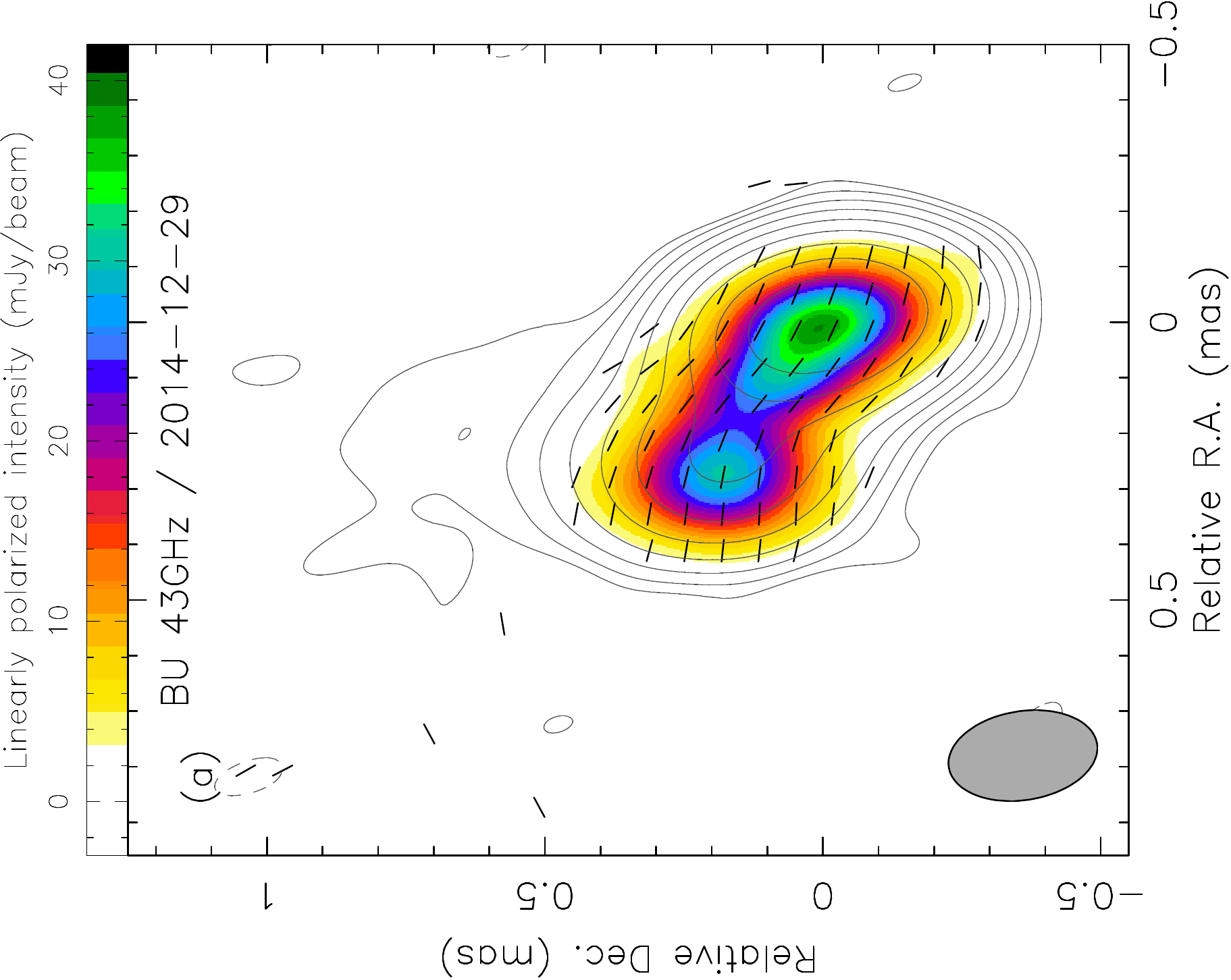}\quad 
\includegraphics[angle=270,width=0.29\textwidth]{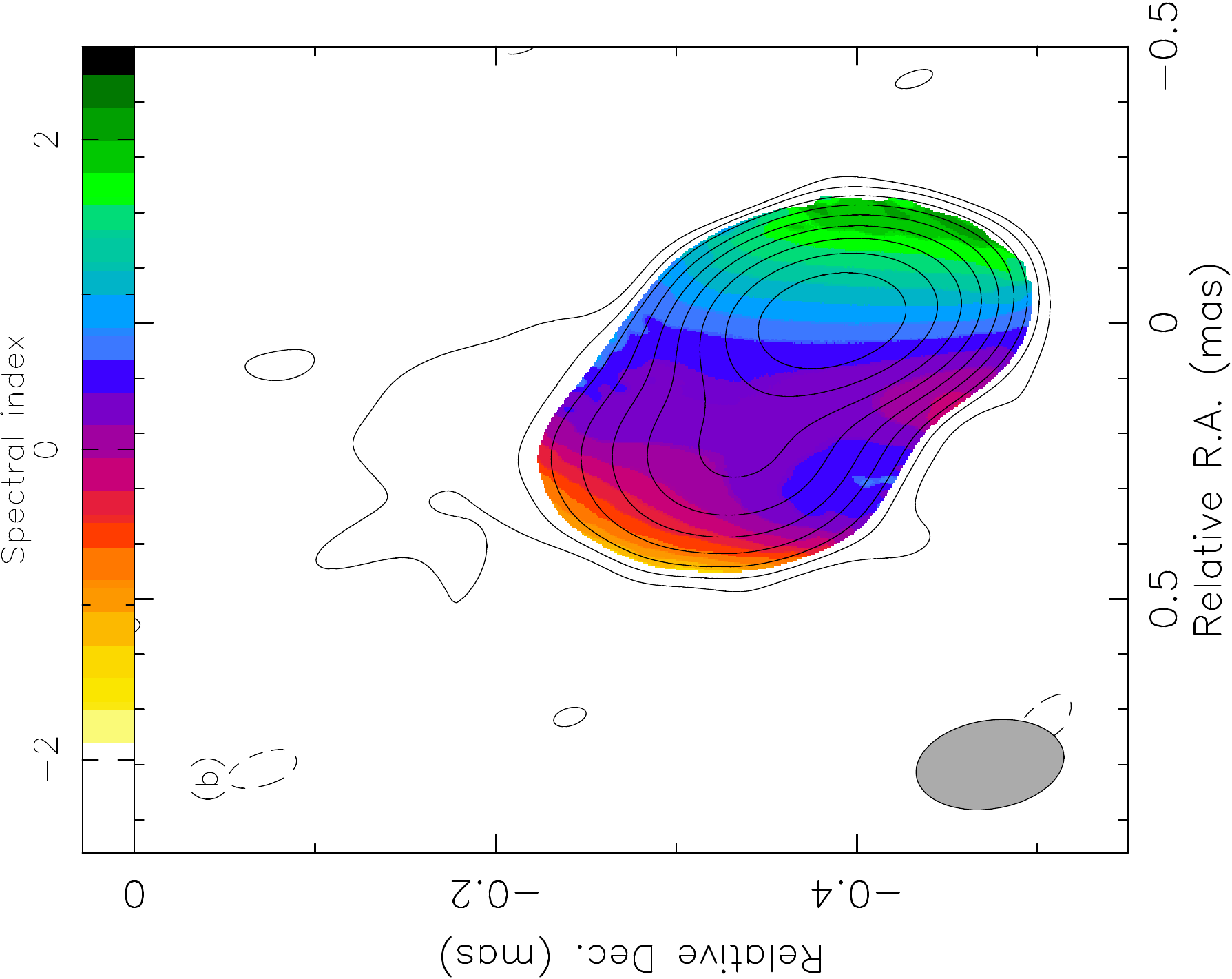}\quad 
\includegraphics[angle=270,width=0.29\textwidth]{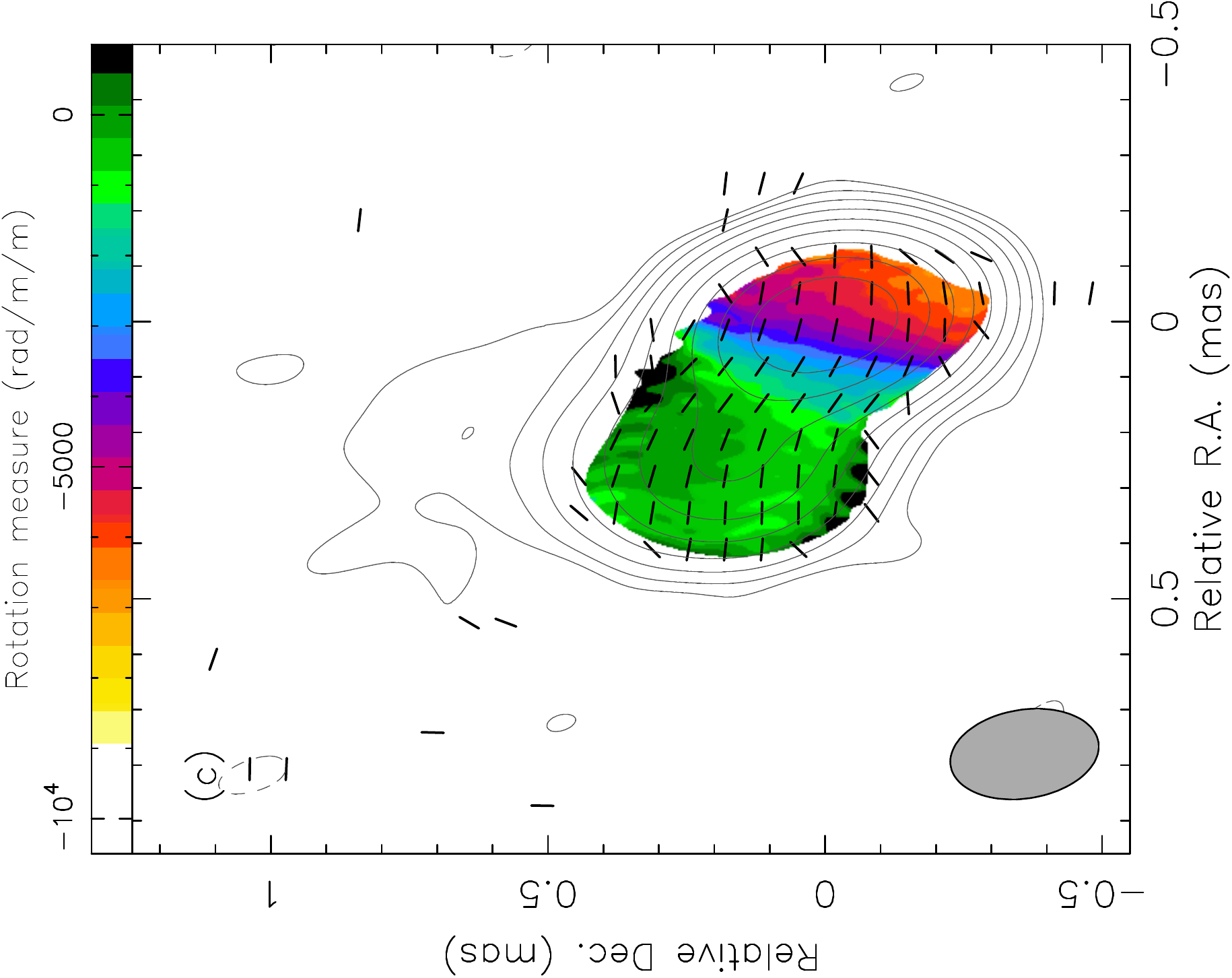} 
\caption{(a) VLBA-BU-BLAZAR image of 0716$+$714 at 43\,GHz obtained at 2014 December 29. The total intensity contours are plotted at $-$2.3, 2.3, 4.6, 9.3, 18.6, 37.2, 74.3, 148.6, 297.3, 594.5, 1189.0~Jy/beam. Linearly polarized intensity is shown in colors starting at 2.1~mJy/beam, and bars indicate the EVPA. Synthesized beam FWHM is 0.27$\times$0.16~mas at a position angle of 9\degr. (b) Spectral index image between the 22\,GHz {\em RadioAstron} and 43\,GHz VLBA-BU-BLAZAR total intensity images, convolved with the 43\,GHz beam, and overlaid on Stokes $I$ contours at 43\,GHz. (c) Rotation Measure map combining the images at 22\,GHz and 43\,GHz, and overlaid on Stokes $I$ contours at 43\,GHz. Color indicates the rotation measure in rad$\cdot \mathrm{m}^{-2}$, and ticks the Faraday-corrected EVPAs at 43\,GHz. 
\label{fig:frm}}
\end{figure*}

As was stated above, the quality of the 15 and 43\,GHz VLBA data taken during the RadioAstron cooling gaps was not good enough for carrying out Faraday rotation measure (RM) analysis, since four antennas of the array were lost during our observations.
Moreover, no MOJAVE observations of 0716+714 were performed close to our epoch.
This limited our analysis to two frequencies: our 22\,GHz data and the BU-BLAZAR 43\,GHz data at a close epoch (2014-12-29, see Fig.~\ref{fig:frm}).
A Faraday rotation analysis based solely on two-frequency data can be performed under the assumption of negligible internal Faraday rotation, in which case we could expect a linear dependence of the polarization vector with the square of the observing frequency. To account for the $n\pi$ ambiguity in the polarization vector we further assume the smallest Faraday rotation. We also note that higher values of $n$ would result in large rotation measure values, and therefore significant bandwidth depolarization, which is not observed. Finally, the rotation measure images of Fig.~\ref{fig:frm} are also performed under the assumption that the polarization structure is not changed in the time lapse between our two frequency observations, and that any differences between polarization angles at the two observed frequencies are due to Faraday rotation.

According to the BU-BLAZAR database, the EVPA in the core region of 0716$+$714 is fairly stable from the end of December 2014 until April 2015, and changes are within 8\degr. 
Likely there was a polarized feature with a reasonably stable EVPA which dominated the linearly polarized intensity in the core area, as we see in our SVLBI image for component C2 (Fig.~\ref{fig:imspace}).  
Thus it is unlikely that polarization properties significantly change over the five-day separation between our observations at 22\,GHz and BU-BLAZAR observations at 43\,GHz on 29 December 2014.

The 22\,GHz {\em RadioAstron} image was convolved with the 43\,GHz VLBA restoring beam size of $0.27\times0.16~$mas at 9$^{\circ}$.
We obtained a shift to the southwest of 0.018~mas to align 22\,GHz image with respect to the 43\,GHz one.
No significant shift of the core position (coreshift) has been detected between these two frequencies, either by referencing the core position to the nearest optically thin component or by using a 2D cross-correlation analysis \citep[see e.g.][]{2000ApJ...530..233W}.
The spectral index (defined as $S\propto\nu^{\alpha}$, where $S$ is the flux density at an observing frequency $\nu$) between 22\,GHz {\em RadioAstron} and 43\,GHz VLBA images is shown in Fig.~\ref{fig:frm}, together with the resultant Faraday RM map.

The RM in the jet is consistent with zero value, whereas the core shows RM of $-(5900\pm1100)~\mathrm{rad~m}^{-2}$.
\citet{2016AA...592L..10L} in their study measured RM between 22\,GHz and 43\,GHz ranging between $-9200$ to 6300 rad m$^{-2}$, which is consistent with our estimates.
The RM measured in the core can be well reconciled with an external Faraday screen located near the jet. 
It is expected that adiabatic losses should result in the decaying particle density and magnetic field along the jet and in the Faraday screen around it, which can explain the higher RM observed in the core of 0716$+$714.
The degree of linear polarization increases from few per cent in the core up to 50 per cent at the position of component C4, Fig.~\ref{fig:imspace}.
Together with the RM gradient along the jet, it may imply significant depolarization near the core region.
The intrinsic EVPA, corrected for the Faraday rotation (Fig.~\ref{fig:frm}) is consistent with the dominance of the toroidal component of the ordered magnetic field in the blazar jet.

\subsection{Timescale of variability}

Following \citet{2017ApJ...846...98J}, the shortest variability timescale $\tau$ expected for component C2 can be calculated using light travel arguments as
\begin{equation}
\tau \thicksim \frac{25.3 RD_\mathrm{L}}{\delta (1+z)},
\label{eq:tau}
\end{equation}
where $D_\mathrm{L}$ is the luminosity distance given in Gpc, $R$ is the FWHM of the circular Gaussian derived from the model fitting in mas, and $\tau$ is in years.
Using the estimated component's size of $R=19~\mu$as (see Table~\ref{tab:mdf}) and the largest observed Doppler factor of $\thicksim25$ \citep{2005AA...433..815B,2017ApJ...846...98J}, we obtain $\tau\thicksim8.6$ days for C2. 
Considering the resolution limit of $R<5~\mu$as \citep{2005astro.ph..3225L}, we obtain $\tau\leq2.3$ days for the unresolved core.
Assuming intrinsic origin of the IDV observed in 0716$+$714, these components explain the total and polarized flux density variations observed in it on scales from a day to a week.
Presence of a more compact component is required to describe the faster, hour-scale variability also reported in 0716$+$714 \citep[e.g.][]{0716Nature,2006AA...452...83B,2015ApJ...809L..27B}.

\subsection{Jet orientation and flaring activity in 0716+714}
\label{s:orient}

\begin{figure*}
\centering
\includegraphics[angle=0,width=0.61\textwidth]{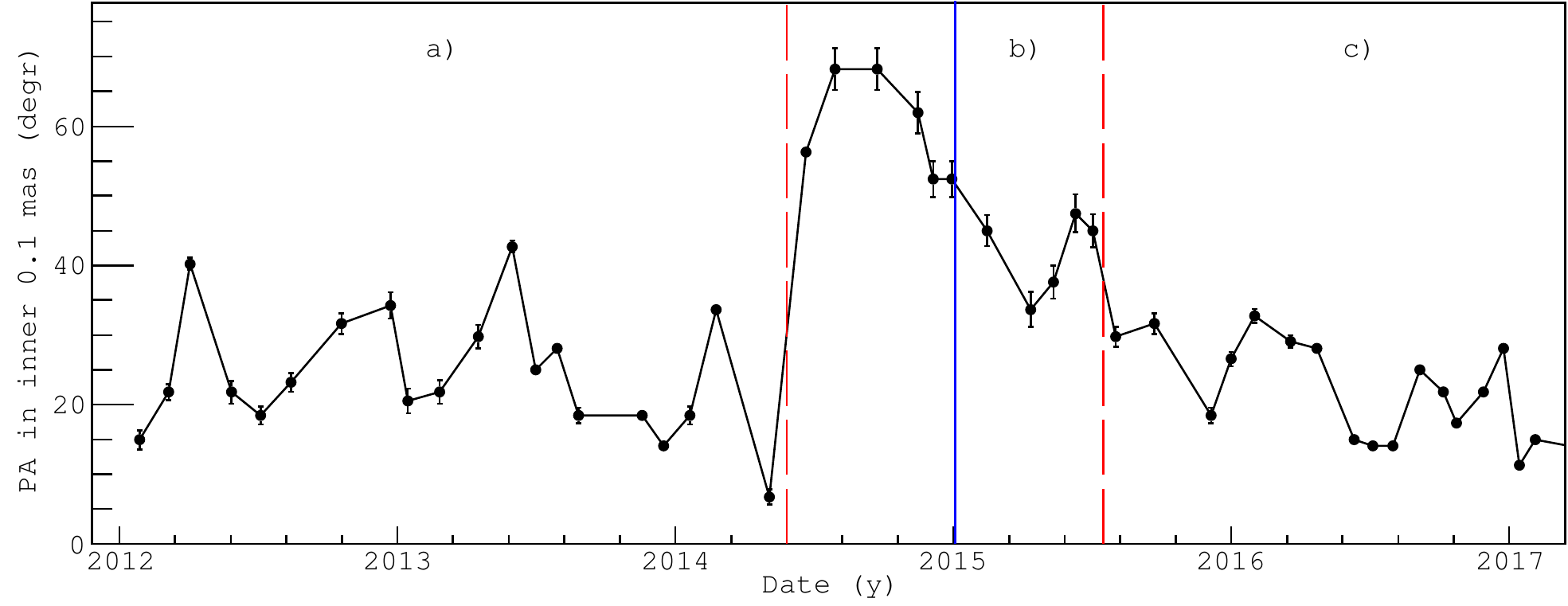}\\
\includegraphics[width=0.61\textwidth]{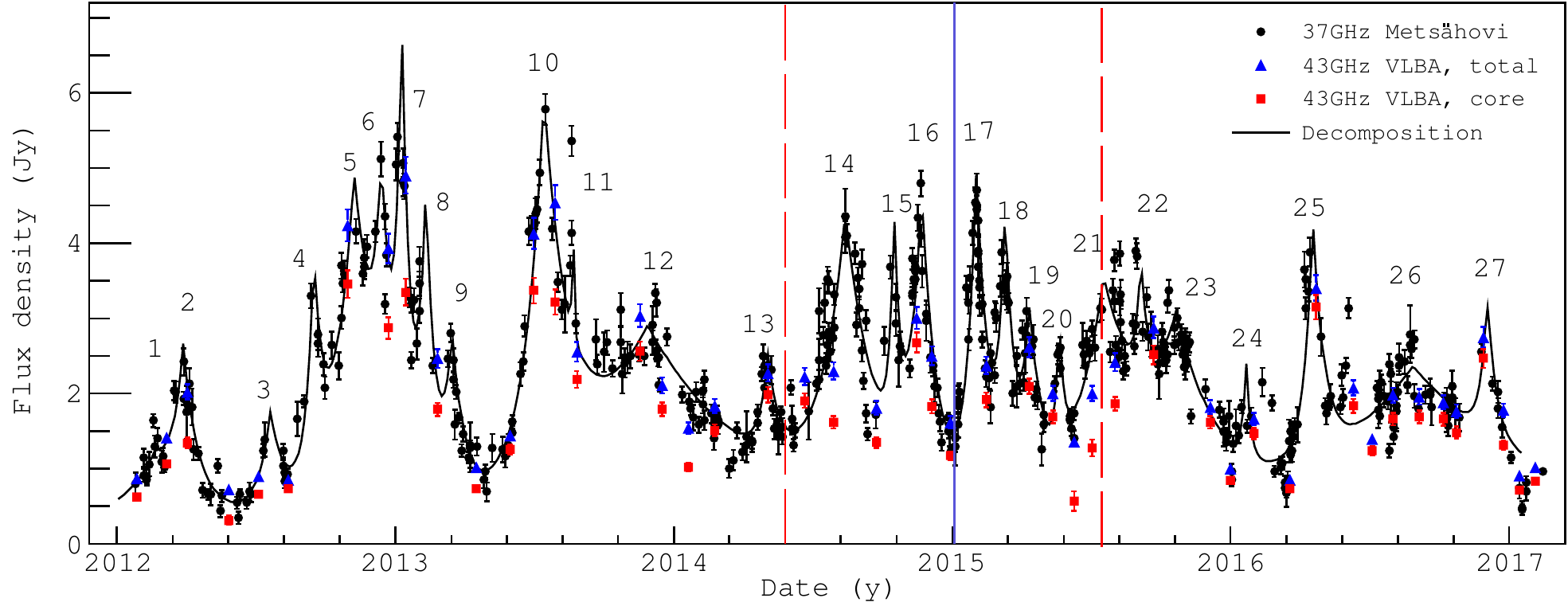}\\
\includegraphics[width=0.61\textwidth]{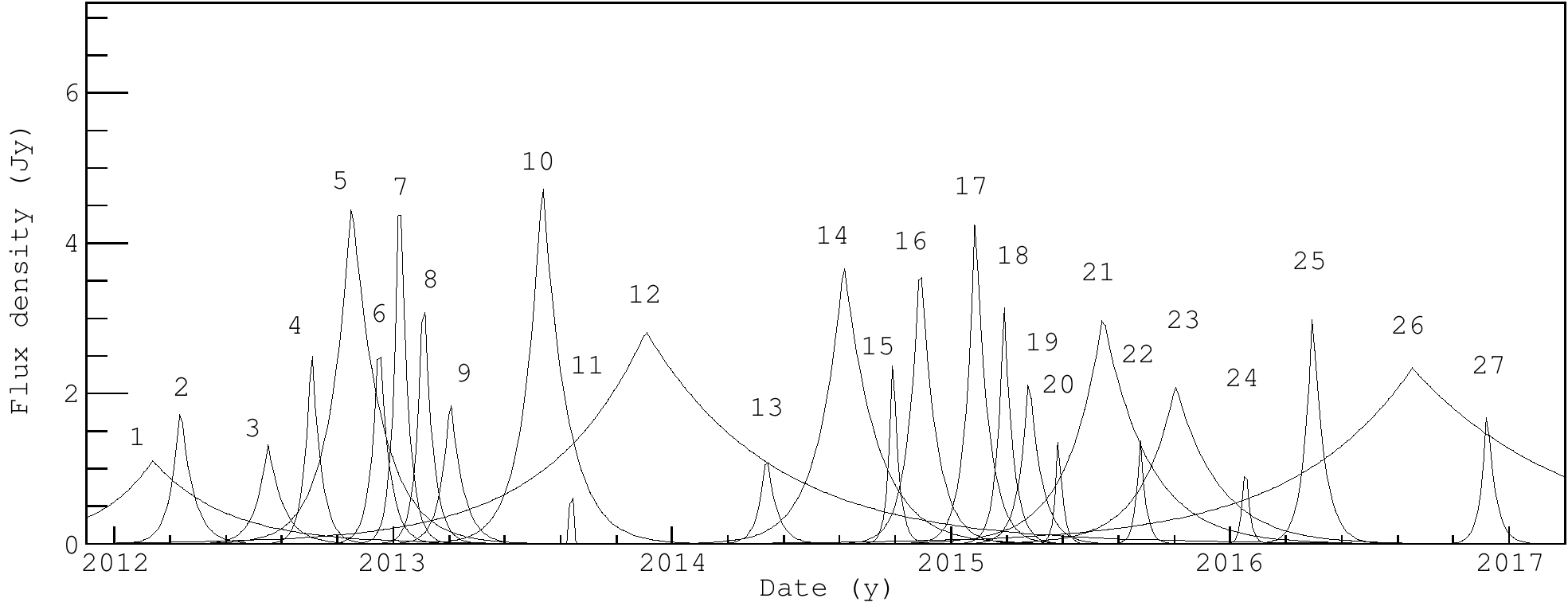}
\caption{(top) The jet PA in the inner 100~$\mu$as at 43\,GHz.
Vertical red dashed lines and letters mark three time ranges, which were used to construct stack images, given in Fig.~\ref{fig:bu_st_im}. Vertical solid blue line indicates the epoch of \textit{RadioAstron} observations.
(middle) 0716$+$714 total flux density at 37\,GHz (black circles, Mets\"{a}hovi), integrated flux density on the VLBA 43\,GHz image (blue triangles) and core flux density at 43\,GHz (red squares). Black solid line represents the results of the decomposition, which is a sum of individual exponential model flares shown at the bottom plot.
\label{fig:tb}}
\end{figure*}

In order to study the evolution of the jet orientation and its potential relation with the jet emission we combine the information about the position angle of the inner jet and the flaring activity in the source during the time period of 2012--2017 (see Fig.~\ref{fig:tb}).

The PA changes are obtained from analysis of the 50 epochs made within VLBA-BU-BLAZAR program at 43\,GHz.
To compare different epochs, we convolve all images with the same beam size of 0.23$\times$0.16~mas at 0\degr, which corresponds to the average beam size over these epochs.
To describe the jet PA we construct ridgelines for each epoch and calculate its direction in the inner $100~\mu$as from the core.
The ridgelines were obtained as the weighted average of the jet emission at a given radial distance from the core, following \citet{2017MNRAS.468.4992P}. 
The algorithm is based on azimuthal slicing of the source image, centered on the core. 
It searches for the weighted average along the slice, i.e. the point where the intensity integrated along the arc is equal on the two sides, using pixels with $\mathrm{SNR}>3$.
The procedure repeats for each slice down the jet.
The final ridgeline is constructed by fitting a cubic spline.
The subtracted PAs are given in Fig.~\ref{fig:tb}.

The flaring activity was investigated using the measurements from the Mets\"{a}hovi 13.7-m dish monitoring program at 37\,GHz and the VLBA-BU-BLAZAR program at 43\,GHz.
The radio light curves at these frequencies in Fig.~\ref{fig:tb} show two large flares at the beginning and mid 2013, followed by other minor outbursts. 
Our \textit{RadioAstron} observations were taken in between some of these minor flares, when it flux density at 37\,GHz reached $\thicksim1.4$\,Jy. 
This is thrice the flux density registered at the absolute minimum ($\thicksim0.4$\,Jy, June 2012), and a quarter of the absolute maximum (5.8\,Jy, July 2013) during 2012--2017.

Following \citet{1999ApJS..120...95V} and \citet{1999ApJ...511..112L}, the flux density variations can be decomposed into individual flares, providing estimates on their physical parameters. 
In order to decompose the observed total flux density variations, we modelled the 37\,GHz Mets\"{a}hovi light curve by a number of individual flares with the following profile
\begin{equation}
S(t) = \left\{
     \begin{array}{lr}
       S_\mathrm{max}e^{(t-t_\mathrm{max})/\tau_\mathrm{var}},\qquad t < t_\mathrm{max},\\
       S_\mathrm{max}e^{(t_\mathrm{max}-t)/1.3\tau_\mathrm{var}},\quad t > t_\mathrm{max},
     \end{array}
   \right.
\end{equation}
where $S_\mathrm{max}$ is the maximum amplitude of the flare, $t_\mathrm{max}$ is the 
epoch of the flare maximum and $\tau_\mathrm{var}$ is the flare rise timescale. 
We identified and fit 27 individual flares, and provide the results of their decomposition in Fig.~\ref{fig:tb} and in Table~\ref{tab:decomp}.

As can be seen from Fig.~\ref{fig:tb}, the jet experiences sudden changes in the PA in the middle of 2014, when its orientation changes from $\sim20\degr$ to $\sim70\degr$ and then returns back to $\sim20\degr$ within a year.
\citet{2013AJ....146..120L} observed periodic variability of the jet PA in 0716$+$714 at larger scales and estimated a variability amplitude of $\sim11\degr$ and a period of $10.9$~yr for it variability.

\begin{figure*}
\includegraphics[angle=0,width=0.28\textwidth]{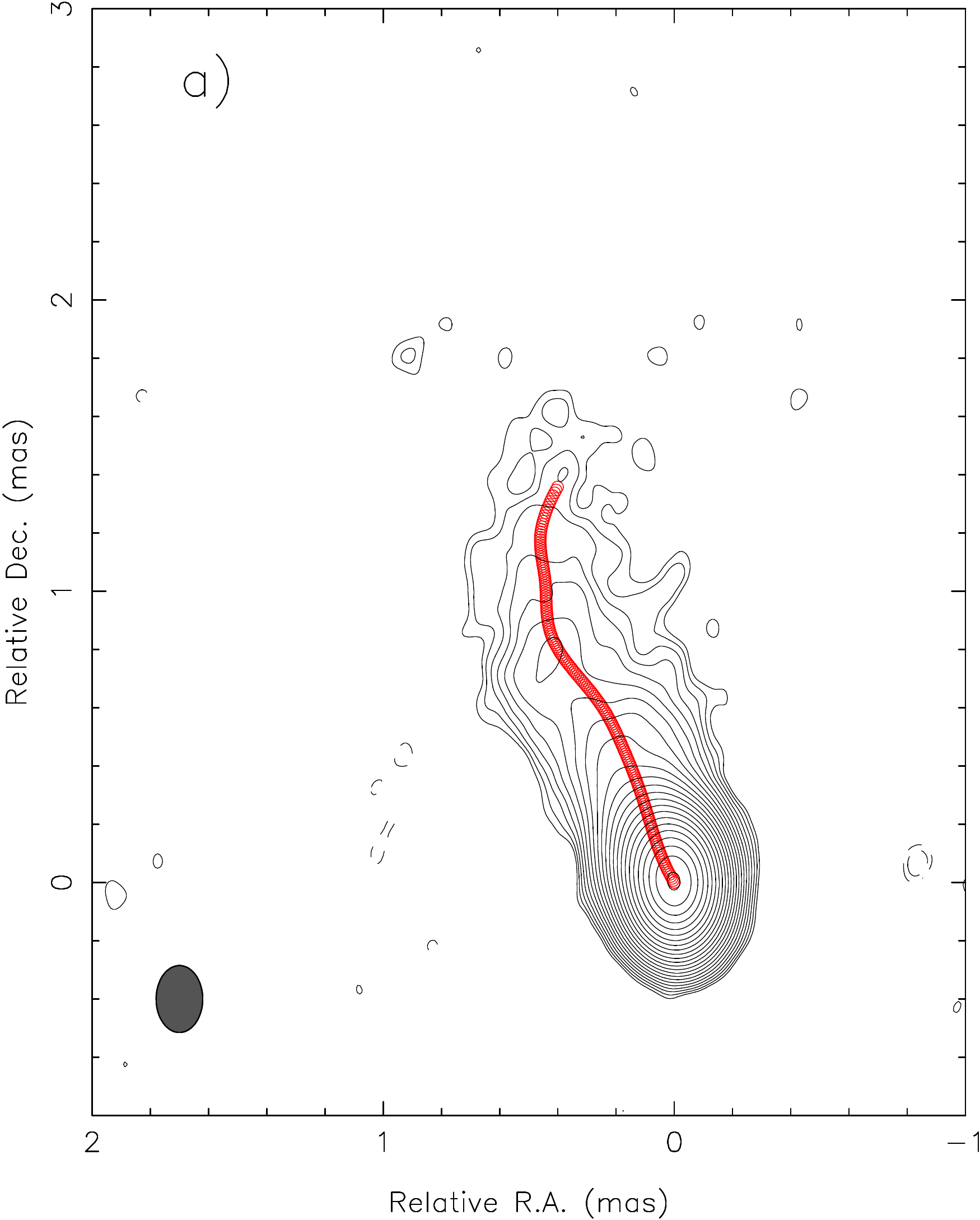}\quad
\includegraphics[angle=0,width=0.28\textwidth]{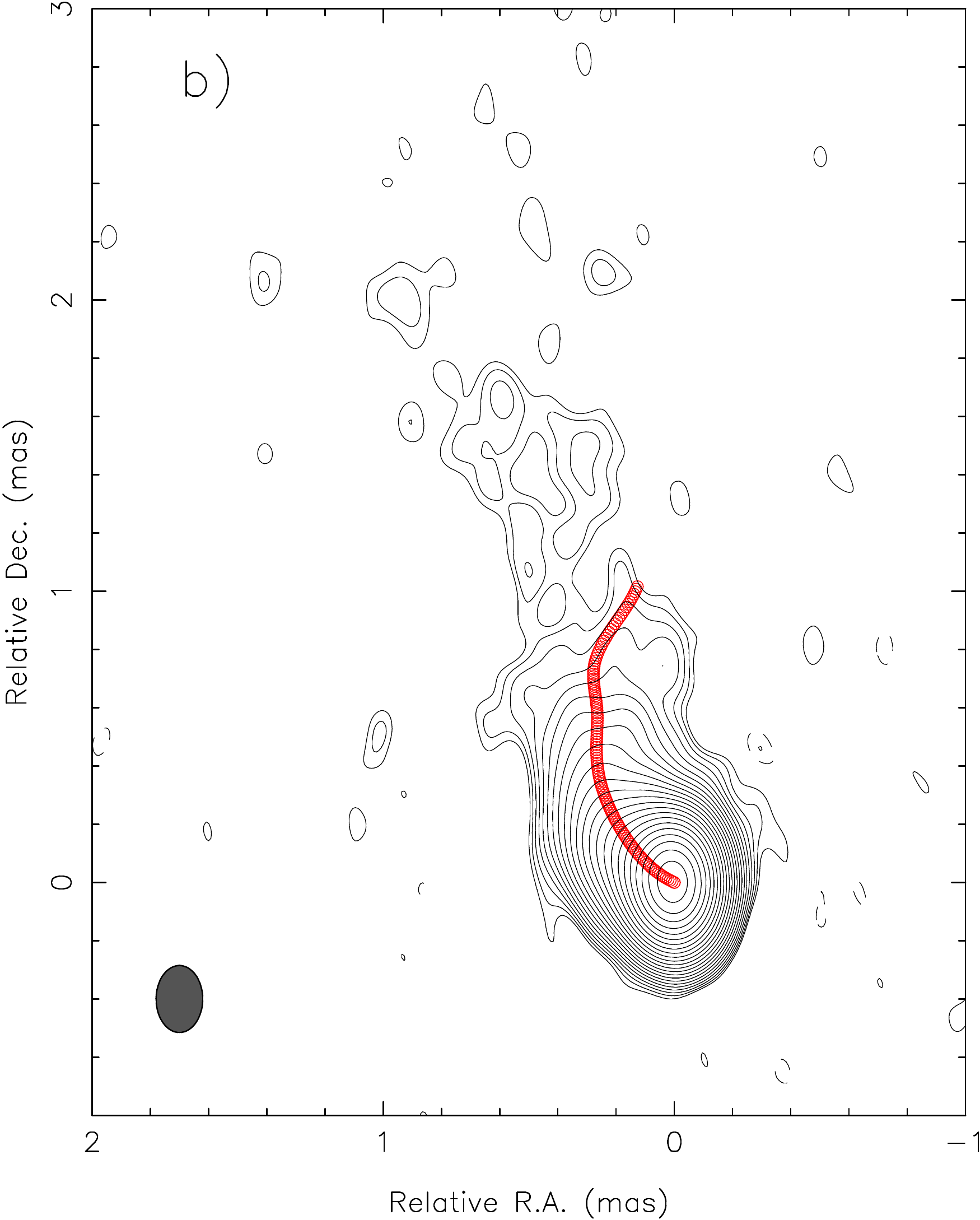}\quad
\includegraphics[angle=0,width=0.28\textwidth]{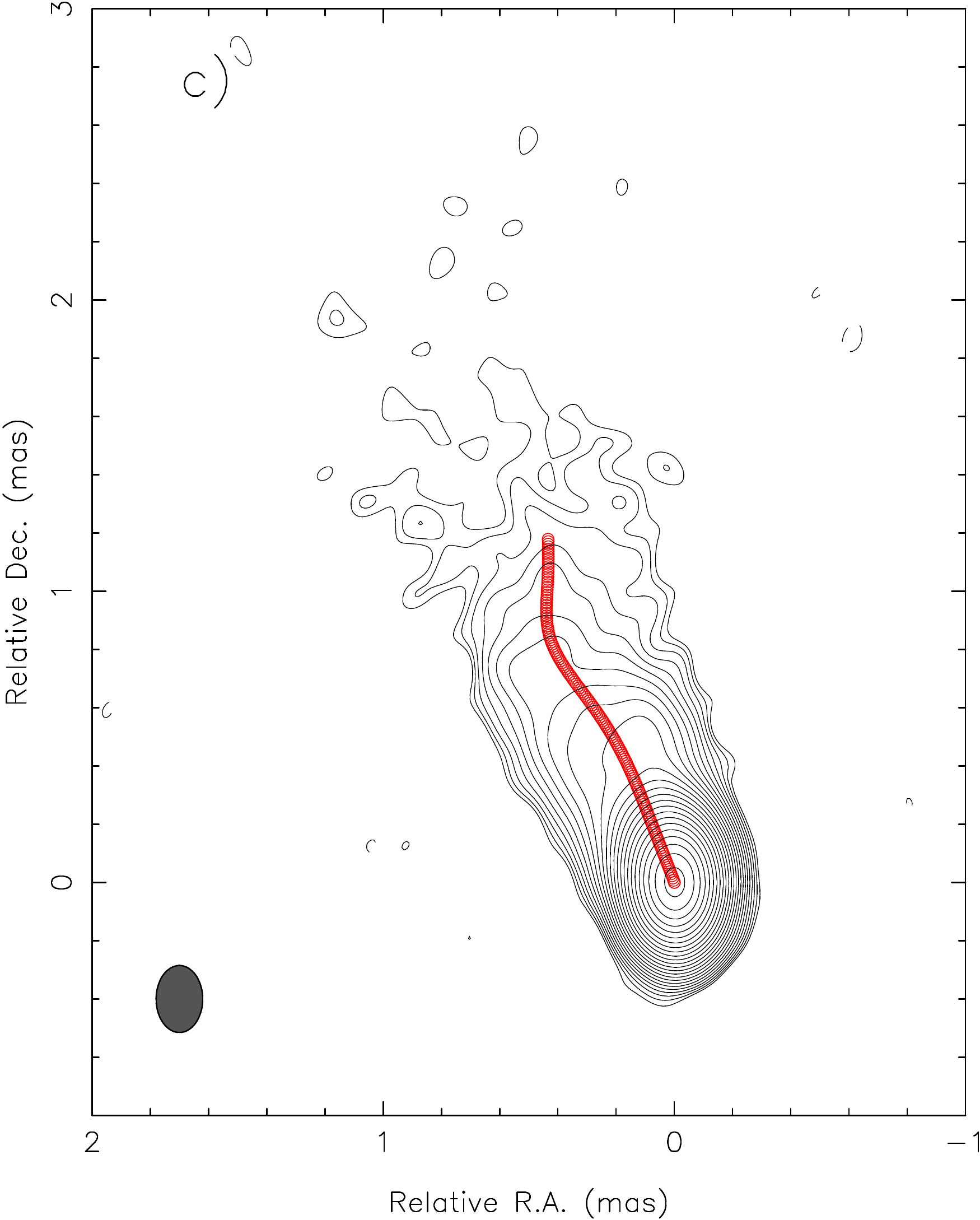}
\caption{Stacked images of 0716$+$714 at 43\,GHz and evolution of the jet shape and ridgeline with time. a) 19 epochs from 2012-01-27 to 2014-02-24. b) 10 epochs from 2014-05-03 to 2015-05-11. c) 21 epochs between 2015-06-09 and 2017-04-16. The ridgeline is given as a number of red circles. The naturally weighted beam is 0.23x0.16 mas at 0 deg and is shown at the left bottom corner.
\label{fig:bu_st_im}}
\end{figure*}

We split the observing epochs into three groups, covering three different time ranges, and isolating the time period when the blazar had the most extreme eastward PA orientation.
These ranges correspond to a) 2012-01-27 -- 2014-02-24, b) 2014-05-03 -- 2015-05-11, and c) 2015-06-09 -- 2017-04-16 (Fig.~\ref{fig:tb}) and are used to construct stacked images of the source presented in Fig.~\ref{fig:bu_st_im}.
As can be seen, our space-VLBI observations probe the source in the middle of the most eastward orientation of the jet.
From Fig.~\ref{fig:tb}, there is no indication for the correlation of variations between the total flux density and the jet PA at scales of 100~$\mu$as from the core.
This suggests, that the variability of the total flux density originates at smaller scales.

\subsection{Variability, brightness temperature, and Doppler boosting}
\label{s:tb_var}

We compute the brightness temperatures $T_\mathrm{b,rf}$ of the model fitted VLBI components in the source rest frame as \citep{2005AJ....130.2473K}
\begin{equation}
T_\mathrm{b,rf} = 1.22\times10^{12}\frac{(1+z)S}{\nu^2\theta_\mathrm{maj}\theta_\mathrm{min}}~[K],
\label{eq:tbgr}
\end{equation}
where $\theta_\mathrm{maj}$ and $\theta_\mathrm{min}$ are the major and minor axes of the Gaussian component in mas, $S$ is the flux density of the component in Jy, and the observing frequency, $\nu$, is given in GHz. 
We can compare brightness temperatures of the model fitted components with the variability brightness temperatures calculated from the flare fits described above. The variability brightness temperature (in the source rest frame), T$_\mathrm{b,var}$, is given by
\begin{equation}
T_\mathrm{b,var} = 1.05\times10^{8} \frac{D_\mathrm{L}^2 S}{(1+z){\nu}^2 \tau_\mathrm{var}^2} ~[K],
\end{equation}
where luminosity distance $D_\mathrm{L}$ is given in Mpc and $\tau_\mathrm{var}$ (yr) is the logarithmic variability timescale, defined as
\begin{equation}
\tau_\mathrm{var} = \frac{\mathrm{d}t}{\mathrm{d}\,\mathrm{ln} S}.
\end{equation}
The variability brightness temperature is related to the variability Doppler factor $\delta_\mathrm{var}$ and the intrinsic brightness temperature by
\begin{equation}
T_\mathrm{b,var} = \delta_\mathrm{var}^3 T_\mathrm{b,int}.
\label{eq:tvar}
\end{equation}
On the other hand, the rest frame $T_\mathrm{b,rf}$ and intrinsic $T_\mathrm{b,int}$ brightness temperatures in the VLBI core are connected through the Doppler boosting factor as
\begin{equation}
T_\mathrm{b,rf} = \delta \, T_\mathrm{b,int}.
\label{eq:tobs}
\end{equation}
Combining equations~\ref{eq:tvar} and \ref{eq:tobs}, it is possible to estimate the Doppler factor and intrinsic brightness temperature.
We use the results of the flare decomposition of the light curve and follow two different methods: (i) assume the value of intrinsic brightness temperature, and (ii) utilize both variability and VLBI data.

\textit{Method~1:}
For $T_\mathrm{b,int}$ we assumed the equipartition value of $5\times10^{10}$\,K \citep{readhead_94}, as discussed in \citet{1999ApJ...511..112L}.
Considering all model fitted flares, it yields the range of Doppler factor values of $3\leq\delta\leq31$ (excluding outlier $\delta\thicksim95$) with a median of $\thicksim18$.
The fastest flares provide better estimate of the Doppler factor, since they suffer less from the smearing with other flares and likely reach limiting brightness temperature.
For the fastest model fitted flares, $\delta$ is $\thicksim30$ and is slightly higher than the maximum value of $\delta\thicksim25$ that has been obtained in other studies of the source \citep[e.g.][]{2005AA...433..815B,2017ApJ...846...98J}.
This may imply that $T_\mathrm{b,int}$ exceeds $T_\mathrm{b,eq}$ during some flares.

\textit{Method~2:}
To calculate rest frame brightness temperature of VLBI jet, we used observations of the blazar performed at 43\,GHz with the VLBA.
We selected the epochs which are observed close in time to the maxima of the decomposed flares (within the $\tau_\mathrm{var}$).
We were able to identify eight flares which meet the criterium.
The respective modelfit parameters of the VLBI core and the estimated $T_\mathrm{b,rf}$, calculated with the equation~\ref{eq:tobs}, are given in Table~\ref{tab:decomp}.
The estimated values of the Doppler factor lie in a range $1\lesssim\delta\lesssim15$, the intrinsic brightness temperature estimates are of $(0.2-2)\times10^{12}$~K (see Table~\ref{tab:decomp}).

\subsection{\textit{RadioAstron} 22~GHz brightness temperature: breaking the Inverse-Compton limit?} 
\label{s:tb_IC}

In case of inverse Compton losses in incoherent synchrotron sources, it is expected that the intrinsic brightness temperature does not exceed the value of $T_\mathrm{b,ic}\approx10^{11.5}$~K \citep{1969ApJ...155L..71K}.
Meanwhile, for the equipartition regime between the energy densities of magnetic field and radiating particles, the value of $T_\mathrm{b,eq}\thicksim5\times10^{10}$~K has been obtained \citep{readhead_94}.

The results of model fitting of \textit{RadioAstron} data yield brightness temperatures of $7\times10^{11}$~K for C1, $1.3\times10^{12}$~K for C2, and $T_\mathrm{b}>2.2\times10^{13}$~K for the core. 
$T_\mathrm{b}$ computed for all the fitted components are given in Table~\ref{tab:mdf}.

The comparison of the Gaussian modelfit results for components C1 and C2 shows that the size of C1 appears to be larger than the size of C2 and that $T_\mathrm{b,C1} < T_\mathrm{b,C2}$. 
Whereas, from the adiabatic expansion and conical jet models (See \citet{2004A&A...426..481K} and references therein) linear dependence of $T_\mathrm{b}$ and size of the components with distance along the jet is expected.
There a number of possibilities that may account for this.
First of all, the linear behavior requires several conditions to be hold: no changes in physical conditions (e.g. absorption) along the strait and continuous jet, constant or slightly varying Doppler factor and the jet viewing angle. 
In Sect.~\ref{s:orient} and \ref{s:tb_var} we show that some of these parameters vary with time.
Change in the collimation profile of the jet from parabolic to a conical one \citep{2012ApJ...745L..28A, 2019arXiv190701485K} will have an impact on this relation.
Blending effects near the core region may be significant, affecting the modelfit results for the size and flux density of the component C1.
Different position angles, at which components C1 and C2 were ejected into jet \citep{2013AJ....146..120L}, and subsequent projection effects will also alter the results.

The brightness temperature in the jet can differ from the estimate, coming from the representation of the structure by a series of Gaussian components.
In this case, the minimum brightness temperature $T_\mathrm{b,min}$ that can be obtained from our data can be directly estimated from the visibility amplitudes, under the condition that structural detail sampled by the given visibility is resolved (See \citealt{2015AA...574A..84L} for further details).
Considering the visibility amplitude on the largest projected baseline between {\em RadioAstron} and Green Bank (of $\thicksim$5.2\,G$\lambda$ on average), the observed brightness temperature of the most compact structure must exceed $9\times10^{12}$~K.
This agrees with $T_\mathrm{b}$ obtained from the Gaussian modelfit for the most compact structure in 0716$+$714 jet, the core.

According to the equation~\ref{eq:tobs} and considering the maximum Doppler factor of $\delta\approx25$, reported by the ground-based kinematic studies of 0716$+$714 \citep[e.g.][]{2005AA...433..815B,2017ApJ...846...98J} and consistent with our estimates in Sect.~\ref{s:tb_var}, the \textit{RadioAstron} observations yield $T_\mathrm{b,int}\geq9\times10^{11}$ in the rest frame of the flow.
It also agrees well with our estimate of $T_\mathrm{b,int} = (0.1-2.0) \times 10^{12}$\,K above.
These estimates imply a departure of physical conditions in the inner jet from the equipartition between magnetic field and relativistic particle energy densities, and are also larger than the inverse Compton brightness limit.

\citet{2016ApJ...820L...9K} suggested that some internal phenomenon may be responsible for producing extremely high brightness temperature, over the Compton catastrophe limit.
The findings of \citet{2017AA...604A.111B} show that states when $T_\mathrm{b,int}$ breaks the abovementioned limits may be connected with the activity of the sources.
Indeed, \citet{2017ApJ...846...98J} in their study observed few cases when $T_\mathrm{b,int}$ of the cores exceeds $T_\mathrm{b,eq}$ by a factor of 10 and associated all of them with strong multiwavelength activity of the sources.

Three weeks after our \textit{RadioAstron} observations the \citet{2018AA...619A..45M} detected an unprecedented outbursts in 0716$+$714 at very high energies (VHE), peaking on January 25 and February 14.
The MAGIC study also report the ejection of a new jet component (K14b, in their notation) which passed through the VLBI core about 55 days before the first VHE flare and moved out with a proper motion of (0.51$\pm$0.09) mas~yr$^{-1}$.
\citet{2018AA...619A..45M} interpreted the VHE flares as being produced by the interaction of this moving component with the stationary feature A1 reported by \citet{2015AA...578A.123R} and \citet{2017ApJ...846...98J}, considered to be a recollimation shock. Using the measured proper motion of K14b and back-extrapolating its position to the epoch of our \textit{RadioAstron} observations, we localized it at a distance of $63\pm42~\mu$as from the core.
It applies that either component C1 or C2 may be corresponding to the component K14b associated with the VHE and multiband flaring activity observed in 0716$+$714 at the beginning of 2015.

\subsection{Doppler factor variability?}

In the case of a continuous, extended jet, the apparent flux will be enhanced by at least a factor of $\delta^2$ relative to that in the AGN rest frame \citep{1985ApJ...295..358L}. 
Recent VLBI study at 86\,GHz \citep[$\lambda=3$\,mm,][]{2019AA...622A..92N} reports significant flux density variations in 0716+714.
From the data taken at two different epochs, 2011-05-05 and 2011-10-09 (their Fig.~6), the following can be seen. The intensity of the extended structure in the jet at first epoch is of about 600\,mJy/beam, meanwhile five months later this structure is not visible down to a level of $\thicksim$20\,mJy/beam.
If the variability is solely due to changes in the amount of Doppler boosting, this 30--fold decrease in flux requires about five times change in Doppler factor.
Meanwhile, flux variations obtained at 43\,GHz (Fig.~\ref{fig:tb}) are less pronounced and imply change in $\delta$ by a factor of three.
Our results of the light curve decomposition indeed indicate (Sect.~\ref{s:orient}) Doppler factor variability. 

If we assume a scenario where rapid variations in Doppler factor are partially due to changes in the jet viewing angle and may be associated with the fast variability observed in the source, we can estimate the respective changes of the jet viewing angle.
For the maximum observed apparent velocity $\beta_\mathrm{app}\approx25c$ \citep{2017ApJ...846...98J}, the required minimum Lorentz factor $\Gamma_\mathrm{min}=(1+\beta^2_\mathrm{app})^{1/2}\approx25$, which yields the velocity of the jet components in units of the speed of light $\beta\approx0.9992$.
Using $\beta\,\mathrm{cos}\theta = 1 - (\delta\Gamma)^{-1}$, the Doppler factor variations in a range of $\delta\thicksim5$ to $\delta\thicksim15$ can be explained by the change in the jet orientation from $\theta\thicksim7^{\circ}$ to $\theta\thicksim4^{\circ}$.

\section{Summary} 
\label{sec:sum}

We present space-VLBI observations of the blazar 0716$+$714 made with the{\em RadioAstron} mission at 22\,GHz.
The observations were performed as part of the AGN Polarization KSP on 2015 January 3-4.
The source has been detected on the space-ground baselines up to 5.6~ED (70 833~km) in total and linearly polarized intensity.
We constructed a full-polarimetric images of 0716$+$714 with a restoring beam of 57$\times$24\,$\mu$as.
At this resolution, we find the complex and significantly bent structure of the central 100~$\mu$as of the jet, which can be represented by two Gaussian components of 32~$\mu$as (C1) and 19~$\mu$as (C2) in size and the unresolved core of $<12\times5~\mu$as in extent.
The jet initially extends towards the south-east at a position angle of 153.1\degr\ up to $\thicksim41~\mu$as (C1), followed by a sharp bend of about 95\degr\ towards the north-east at a distance $\thicksim58~\mu$as from the core (C2).
Complementary analysis of the VLBA data at 43\,GHz taken in 2012--2017 shows, that our \textit{RadioAstron} experiment was made close to the time when blazar jet had the most eastward orientation during these five years of observations.
We conclude, that the inner jet of 0716$+$714 appears bent, as it observed at viewing angle which is smaller than the opening angle of the conical outflow.

From the comparison of our {\em RadioAstron} results with the large-scale radio image of 0716$+$714, we see about 210\degr\, offset between orientations of the component C1 and the arcsec-scale structure of the jet relative to the core. 
Due to projection effects, this apparent PA offset may correspond to the intrinsic bend of the jet of about $\thicksim10^{\circ}$. The mechanism capable of producing this curved structure of the blazar, remains open.

The polarization image of 0716$+$714 is characterized by a fairly strong 15\%-linearly-polarized component, located $\sim58~\mu$as downstream from the core, which is represented by the model fitted component of $19~\mu$as in size.
This jet component may be responsible for the intrinsic variability of the blazar total and polarized intensities timescales of about a week, while the unresolved core of the jet can be responsible for variability on timescales down to about two days.

From the joint analysis of our observations together with VLBA data at 43\,GHz, we estimate rotation measure of $-(5900\pm1100)$~rad~m$^{-2}$ in the core. 
We do not detect a significant Faraday rotation in the jet.
We conclude that the external Faraday screen in close vicinity to the jet is the most promising explanation for the observed rotation measure, and that the blazar jet has a well ordered magnetic field with the a dominant toroidal component.

The intrinsic brightness temperature of the jet core is found to be $T_\mathrm{b,int}>9\times10^{11}$~K when assuming the Doppler factor of $\delta\thicksim25$.
This value exceeds the equipartition and inverse Compton limits by factors of 18 and 3, respectively.
From the decomposition of the 37\,GHz light curve into individual flares and ground based VLBA observations, we estimate the Doppler factor values in a range of $1 \lesssim \delta \lesssim 15$, while intrinsic brightness temperature is found to vary between $2\times10^{11}$\,K and $2\times10^{12}$\,K.
We suggest, that variations of $\delta$ and $T_\mathrm{b,int}$ are due to changes in the jet viewing angle.

\acknowledgments
{We thank the anonymous referee for useful comments which helped to improve the manuscript.
The authors are grateful to Uwe Bach for providing measurements of the absolute EVPA orientation obtained at Effelsberg.
EVK acknowledges support from the Italian Space Agency under contract ASI-INAF 2015-023-R.O.
JLG and AF was supported by the Spanish Ministry of Economy and Competitiveness grants AYA2013-40825-P and AYA2016-80889-P.
YYK was supported by the government of the Russian Federation (agreement 05.Y09.21.0018) and the Alexander von Humboldt Foundation.
GB acknowledges financial support under the INTEGRAL ASI-INAF agreement 2013-025.R01.
The \textit{RadioAstron} project is led by the Astro Space Center of the Lebedev Physical Institute of the Russian Academy of Sciences and the Lavochkin Scientific and Production Association under a contract with the State Space Corporation ROSCOSMOS, in collaboration with partner organizations in Russia and other countries.
Results of optical positioning measurements of the Spektr-R spacecraft by the global MASTER Robotic Net \citep{2010AdAst2010E..30L}, ISON collaboration, and Kourovka observatory were used for spacecraft orbit determination in addition to mission facilities.
This research has made use of data from the MOJAVE database that is maintained by the MOJAVE team \citep{2018ApJS..234...12L}.
This study makes use of 43\,GHz VLBA data from the VLBA-BU Blazar Monitoring Program (VLBA-BU-BLAZAR), funded by NASA through the Fermi Guest Investigator grants, the most recent 80NSSC17K0649.
The VLBA is an instrument of the Long Baseline Observatory. 
The Long Baseline Observatory is a facility of the National Science Foundation operated by Associated Universities, Inc. 
This publication makes use of data obtained at the Mets\"{a}hovi Radio Observatory, operated by the Aalto University.
This work is partly based on observations carried out using the 100-m telescope of the MPIfR (Max-Planck-Institute for Radio Astronomy) at Effelsberg, the Noto telescope operated by INAF - Istituto di Radioastronomia and the 32-meter radio telescope operated by Torun Centre for Astronomy of Nicolaus Copernicus University in Torun (Poland) and supported by the Polish Ministry of Science and Higher Education SpUB grant.
The European VLBI Network is a joint facility of independent European, African, Asian, and North American radio astronomy institutes. 
Scientific results from data presented in this publication are derived from the following global VLBI project code: GL041.}

\facilities{{\em RadioAstron} Space Radio Telescope (Spektr-R), VLBA, Green Bank 100-m radio telescope, Sheshan 25-m radio telescope (Shangai), EVN, Effelsberg 100-m radio telescope, Noto 32-m radio telescope}

\software{AIPS \citep{aips}, Difmap \citep{1994AAS...185.0808P}, ROOT framework \citep{ANTCHEVA20092499}}

\bibliographystyle{aasjournal}
\bibliography{evk_ra0716}

\begin{thebibliography}{}
\expandafter\ifx\csname natexlab\endcsname\relax\def\natexlab#1{#1}\fi
\providecommand{\url}[1]{\href{#1}{#1}}
\providecommand{\dodoi}[1]{doi:~\href{http://doi.org/#1}{\nolinkurl{#1}}}
\providecommand{\doeprint}[1]{\href{http://ascl.net/#1}{\nolinkurl{http://ascl.net/#1}}}
\providecommand{\doarXiv}[1]{\href{https://arxiv.org/abs/#1}{\nolinkurl{https://arxiv.org/abs/#1}}}

\bibitem[{{Antcheva} {et~al.}(2009){Antcheva}, {Ballintijn}, {Bellenot},
  {Biskup}, {Brun}, {Buncic}, {Canal}, {Casadei}, {Couet}, {Fine}, {Franco},
  {Ganis}, {Gheata}, {Gonzalez Maline}, {Goto}, {Iwaszkiewicz}, {Kreshuk},
  {Marcos Segura}, {Maunder}, {Moneta}, {Naumann}, {Offermann}, {Onuchin},
  {Panacek}, {Rademakers}, {Russo}, \& {Tadel}}]{ANTCHEVA20092499}
{Antcheva}, I., {Ballintijn}, M., {Bellenot}, B., {et~al.} 2009, Computer
  Physics Communications, 180, 2499 , \dodoi{10.1016/j.cpc.2009.08.005}

\bibitem[{{Asada} \& {Nakamura}(2012)}]{2012ApJ...745L..28A}
{Asada}, K., \& {Nakamura}, M. 2012, \apjl, 745, L28,
  \dodoi{10.1088/2041-8205/745/2/L28}

\bibitem[{{Bach} {et~al.}(2006){Bach}, {Krichbaum}, {Kraus}, {Witzel}, \&
  {Zensus}}]{2006AA...452...83B}
{Bach}, U., {Krichbaum}, T.~P., {Kraus}, A., {Witzel}, A., \& {Zensus}, J.~A.
  2006, \aap, 452, 83, \dodoi{10.1051/0004-6361:20053943}

\bibitem[{{Bach} {et~al.}(2005){Bach}, {Krichbaum}, {Ros}, {Britzen}, {Tian},
  {Kraus}, {Witzel}, \& {Zensus}}]{2005AA...433..815B}
{Bach}, U., {Krichbaum}, T.~P., {Ros}, E., {et~al.} 2005, \aap, 433, 815,
  \dodoi{10.1051/0004-6361:20040388}

\bibitem[{{Bennett} {et~al.}(2014){Bennett}, {Larson}, {Weiland}, \&
  {Hinshaw}}]{2014ApJ...794..135B}
{Bennett}, C.~L., {Larson}, D., {Weiland}, J.~L., \& {Hinshaw}, G. 2014, \apj,
  794, 135, \dodoi{10.1088/0004-637X/794/2/135}

\bibitem[{{Bhatta} {et~al.}(2015){Bhatta}, {Goyal}, {Ostrowski}, {Stawarz},
  {Akitaya}, {Arkharov}, {Bachev}, {Ben{\'{\i}}tez}, {Borman}, {Carosati},
  {Cason}, {Damljanovic}, {Dhalla}, {Frasca}, {Hu}, {Itoh}, {Jorstad},
  {Jableka}, {Kawabata}, {Klimanov}, {Kurtanidze}, {Larionov}, {Laurence},
  {Leto}, {Markowitz}, {Marscher}, {Moody}, {Moritani}, {Ohlert}, {Di Paola},
  {Raiteri}, {Rizzi}, {Sadun}, {Sasada}, {Sergeev}, {Strigachev}, {Takaki},
  {Troitsky}, {Ui}, {Villata}, {Vince}, {Webb}, {Yoshida}, {Zola}, \&
  {Hiriart}}]{2015ApJ...809L..27B}
{Bhatta}, G., {Goyal}, A., {Ostrowski}, M., {et~al.} 2015, \apjl, 809, L27,
  \dodoi{10.1088/2041-8205/809/2/L27}

\bibitem[{{Bruni} {et~al.}(2016){Bruni}, {Anderson}, {Alef}, {Rottmann},
  {Lobanov}, \& {Zensus}}]{2016Galax...4...55B}
{Bruni}, G., {Anderson}, J., {Alef}, W., {et~al.} 2016, Galaxies, 4, 55,
  \dodoi{10.3390/galaxies4040055}

\bibitem[{{Bruni} {et~al.}(2017){Bruni}, {G{\'o}mez}, {Casadio}, {Lobanov},
  {Kovalev}, {Sokolovsky}, {Lisakov}, {Bach}, {Marscher}, {Jorstad},
  {Anderson}, {Krichbaum}, {Savolainen}, {Vega-Garc{\'{\i}}a}, {Fuentes},
  {Zensus}, {Alberdi}, {Lee}, {Lu}, {P{\'e}rez-Torres}, \&
  {Ros}}]{2017AA...604A.111B}
{Bruni}, G., {G{\'o}mez}, J.~L., {Casadio}, C., {et~al.} 2017, \aap, 604, A111,
  \dodoi{10.1051/0004-6361/201731220}

\bibitem[{{Danforth} {et~al.}(2013){Danforth}, {Nalewajko}, {France}, \&
  {Keeney}}]{2013ApJ...764...57D}
{Danforth}, C.~W., {Nalewajko}, K., {France}, K., \& {Keeney}, B.~A. 2013,
  \apj, 764, 57, \dodoi{10.1088/0004-637X/764/1/57}

\bibitem[{{Event Horizon Telescope Collaboration}
  {et~al.}(2019{\natexlab{a}}){Event Horizon Telescope Collaboration},
  {Akiyama}, {Alberdi}, {Alef}, {Asada}, {Azulay}, {Baczko}, {Ball},
  {Balokovi{\'c}}, \& {Barrett}}]{2019ApJ...875L...4E}
{Event Horizon Telescope Collaboration}, {Akiyama}, K., {Alberdi}, A., {et~al.}
  2019{\natexlab{a}}, \apjl, 875, L4, \dodoi{10.3847/2041-8213/ab0e85}

\bibitem[{{Event Horizon Telescope Collaboration}
  {et~al.}(2019{\natexlab{b}}){Event Horizon Telescope Collaboration},
  {Akiyama}, {Alberdi}, {Alef}, {Asada}, {Azulay}, {Baczko}, {Ball},
  {Balokovi{\'c}}, {Barrett}, {Bintley}, {Blackburn}, {Boland}, {Bouman},
  {Bower}, {Bremer}, {Brinkerink}, {Brissenden}, {Britzen}, {Broderick},
  {Broguiere}, {Bronzwaer}, {Byun}, {Carlstrom}, {Chael}, {Chan}, {Chatterjee},
  {Chatterjee}, {Chen}, {Chen}, {Cho}, {Christian}, {Conway}, {Cordes}, {Crew},
  {Cui}, {Davelaar}, {De Laurentis}, {Deane}, {Dempsey}, {Desvignes}, {Dexter},
  {Doeleman}, {Eatough}, {Falcke}, {Fish}, {Fomalont}, {Fraga-Encinas},
  {Friberg}, {Fromm}, {G{\'o}mez}, {Galison}, {Gammie}, {Garc{\'\i}a},
  {Gentaz}, {Georgiev}, {Goddi}, {Gold}, {Gu}, {Gurwell}, {Hada}, {Hecht},
  {Hesper}, {Ho}, {Ho}, {Honma}, {Huang}, {Huang}, {Hughes}, {Ikeda}, {Inoue},
  {Issaoun}, {James}, {Jannuzi}, {Janssen}, {Jeter}, {Jiang}, {Johnson},
  {Jorstad}, {Jung}, {Karami}, {Karuppusamy}, {Kawashima}, {Keating},
  {Kettenis}, {Kim}, {Kim}, {Kim}, {Kino}, {Koay}, {Koch}, {Koyama}, {Kramer},
  {Kramer}, {Krichbaum}, {Kuo}, {Lauer}, {Lee}, {Li}, {Li}, {Lindqvist}, {Liu},
  {Liuzzo}, {Lo}, {Lobanov}, {Loinard}, {Lonsdale}, {Lu}, {MacDonald}, {Mao},
  {Markoff}, {Marrone}, {Marscher}, {Mart{\'\i}-Vidal}, {Matsushita},
  {Matthews}, {Medeiros}, {Menten}, {Mizuno}, {Mizuno}, {Moran}, {Moriyama},
  {Moscibrodzka}, {M{\"u}ller}, {Nagai}, {Nagar}, {Nakamura}, {Narayan},
  {Narayanan}, {Natarajan}, {Neri}, {Ni}, {Noutsos}, {Okino}, {Olivares},
  {Ortiz-Le{\'o}n}, {Oyama}, {{\"O}zel}, {Palumbo}, {Patel}, {Pen}, {Pesce},
  {Pi{\'e}tu}, {Plambeck}, {PopStefanija}, {Porth}, {Prather},
  {Preciado-L{\'o}pez}, {Psaltis}, {Pu}, {Ramakrishnan}, {Rao}, {Rawlings},
  {Raymond}, {Rezzolla}, {Ripperda}, {Roelofs}, {Rogers}, {Ros}, {Rose},
  {Roshanineshat}, {Rottmann}, {Roy}, {Ruszczyk}, {Ryan}, {Rygl},
  {S{\'a}nchez}, {S{\'a}nchez-Arguelles}, {Sasada}, {Savolainen}, {Schloerb},
  {Schuster}, {Shao}, {Shen}, {Small}, {Sohn}, {SooHoo}, {Tazaki}, {Tiede},
  {Tilanus}, {Titus}, {Toma}, {Torne}, {Trent}, {Trippe}, {Tsuda}, {van
  Bemmel}, {van Langevelde}, {van Rossum}, {Wagner}, {Wardle}, {Weintroub},
  {Wex}, {Wharton}, {Wielgus}, {Wong}, {Wu}, {Young}, {Young}, {Younsi},
  {Yuan}, {Yuan}, {Zensus}, {Zhao}, {Zhao}, {Zhu}, {Algaba}, {Allardi},
  {Amestica}, {Bach}, {Beaudoin}, {Benson}, {Berthold}, {Blanchard},
  {Blundell}, {Bustamente}, {Cappallo}, {Castillo-Dom{\'\i}nguez}, {Chang},
  {Chang}, {Chang}, {Chen}, {Chilson}, {Chuter}, {C{\'o}rdova Rosado},
  {Coulson}, {Crawford}, {Crowley}, {David}, {Derome}, {Dexter}, {Dornbusch},
  {Dudevoir}, {Dzib}, {Eckert}, {Erickson}, {Everett}, {Faber}, {Farah},
  {Fath}, {Folkers}, {Forbes}, {Freund}, {G{\'o}mez-Ruiz}, {Gale}, {Gao},
  {Geertsema}, {Graham}, {Greer}, {Grosslein}, {Gueth}, {Halverson}, {Han},
  {Han}, {Hao}, {Hasegawa}, {Henning}, {Hern{\'a}ndez-G{\'o}mez},
  {Herrero-Illana}, {Heyminck}, {Hirota}, {Hoge}, {Huang}, {Impellizzeri},
  {Jiang}, {Kamble}, {Keisler}, {Kimura}, {Kono}, {Kubo}, {Kuroda}, {Lacasse},
  {Laing}, {Leitch}, {Li}, {Lin}, {Liu}, {Liu}, {Lu}, {Marson},
  {Martin-Cocher}, {Massingill}, {Matulonis}, {McColl}, {McWhirter}, {Messias},
  {Meyer-Zhao}, {Michalik}, {Monta{\~n}a}, {Montgomerie}, {Mora-Klein},
  {Muders}, {Nadolski}, {Navarro}, {Nguyen}, {Nishioka}, {Norton}, {Nystrom},
  {Ogawa}, {Oshiro}, {Oyama}, {Padin}, {Parsons}, {Paine}, {Pe{\~n}alver},
  {Phillips}, {Poirier}, {Pradel}, {Primiani}, {Raffin}, {Rahlin}, {Reiland},
  {Risacher}, {Ruiz}, {S{\'a}ez-Mada{\'\i}n}, {Sassella}, {Schellart}, {Shaw},
  {Silva}, {Shiokawa}, {Smith}, {Snow}, {Souccar}, {Sousa}, {Sridharan},
  {Srinivasan}, {Stahm}, {Stark}, {Story}, {Timmer}, {Vertatschitsch},
  {Walther}, {Wei}, {Whitehorn}, {Whitney}, {Woody}, {Wouterloot}, {Wright},
  {Yamaguchi}, {Yu}, {Zeballos}, \& {Ziurys}}]{2019ApJ...875L...2E}
---. 2019{\natexlab{b}}, \apjl, 875, L2, \dodoi{10.3847/2041-8213/ab0c96}

\bibitem[{{Fuhrmann} {et~al.}(2008){Fuhrmann}, {Krichbaum}, {Witzel}, {Kraus},
  {Britzen}, {Bernhart}, {Impellizzeri}, {Agudo}, {Klare}, {Sohn}, {Angelakis},
  {Bach}, {Gab{\'a}nyi}, {K{\"o}rding}, {Pagels}, {Zensus}, {Wagner},
  {Ostorero}, {Ungerechts}, {Grewing}, {Tornikoski}, {Apponi},
  {Vila-Vilar{\'o}}, {Ziurys}, \& {Strom}}]{2008AA...490.1019F}
{Fuhrmann}, L., {Krichbaum}, T.~P., {Witzel}, A., {et~al.} 2008, \aap, 490,
  1019, \dodoi{10.1051/0004-6361:20078893}

\bibitem[{{Gabuzda} {et~al.}(2000){Gabuzda}, {Kochenov}, {Cawthorne}, \&
  {Kollgaard}}]{2000MNRAS.313..627G}
{Gabuzda}, D.~C., {Kochenov}, P.~Y., {Cawthorne}, T.~V., \& {Kollgaard}, R.~I.
  2000, \mnras, 313, 627, \dodoi{10.1046/j.1365-8711.2000.03309.x}

\bibitem[{{G{\'o}mez} {et~al.}(2016){G{\'o}mez}, {Lobanov}, {Bruni}, {Kovalev},
  {Marscher}, {Jorstad}, {Mizuno}, {Bach}, {Sokolovsky}, {Anderson}, {Galindo},
  {Kardashev}, \& {Lisakov}}]{2016ApJ...817...96G}
{G{\'o}mez}, J.~L., {Lobanov}, A.~P., {Bruni}, G., {et~al.} 2016, \apj, 817,
  96, \dodoi{10.3847/0004-637X/817/2/96}

\bibitem[{{Greisen}(2003)}]{aips}
{Greisen}, E.~W. 2003, in Astrophysics and Space Science Library 285,
  Information Handling in Astronomy -- Historical Vistas, ed. A.~{Heck}
  (Dordrecht: Kluwer), 109

\bibitem[{{Gupta} {et~al.}(2012){Gupta}, {Krichbaum}, {Wiita}, {Rani},
  {Sokolovsky}, {Mohan}, {Mangalam}, {Marchili}, {Fuhrmann}, {Agudo}, {Bach},
  {Bachev}, {B{\"o}ttcher}, {Gabanyi}, {Gaur}, {Hawkins}, {Kimeridze},
  {Kurtanidze}, {Kurtanidze}, {Lee}, {Liu}, {McBreen}, {Nesci}, {Nestoras},
  {Nikolashvili}, {Ohlert}, {Palma}, {Peneva}, {Pursimo}, {Semkov},
  {Strigachev}, {Webb}, {Wiesemeyer}, \& {Zensus}}]{2012MNRAS.425.1357G}
{Gupta}, A.~C., {Krichbaum}, T.~P., {Wiita}, P.~J., {et~al.} 2012, \mnras, 425,
  1357, \dodoi{10.1111/j.1365-2966.2012.21550.x}

\bibitem[{{Hovatta} {et~al.}(2009){Hovatta}, {Valtaoja}, {Tornikoski}, \&
  {L{\"a}hteenm{\"a}ki}}]{2009AA...494..527H}
{Hovatta}, T., {Valtaoja}, E., {Tornikoski}, M., \& {L{\"a}hteenm{\"a}ki}, A.
  2009, \aap, 494, 527, \dodoi{10.1051/0004-6361:200811150}

\bibitem[{{Hu} {et~al.}(2014){Hu}, {Chen}, {Guo}, {Jiang}, \&
  {Li}}]{2014MNRAS.443.2940H}
{Hu}, S.~M., {Chen}, X., {Guo}, D.~F., {Jiang}, Y.~G., \& {Li}, K. 2014,
  \mnras, 443, 2940, \dodoi{10.1093/mnras/stu1373}

\bibitem[{{Jennison}(1958)}]{1958MNRAS.118..276J}
{Jennison}, R.~C. 1958, \mnras, 118, 276, \dodoi{10.1093/mnras/118.3.276}

\bibitem[{{Jorstad} {et~al.}(2017){Jorstad}, {Marscher}, {Morozova},
  {Troitsky}, {Agudo}, {Casadio}, {Foord}, {G{\'o}mez}, {MacDonald}, {Molina},
  {L{\"a}hteenm{\"a}ki}, {Tammi}, \& {Tornikoski}}]{2017ApJ...846...98J}
{Jorstad}, S.~G., {Marscher}, A.~P., {Morozova}, D.~A., {et~al.} 2017, \apj,
  846, 98, \dodoi{10.3847/1538-4357/aa8407}

\bibitem[{{Kadler} {et~al.}(2004){Kadler}, {Ros}, {Lobanov}, {Falcke}, \&
  {Zensus}}]{2004A&A...426..481K}
{Kadler}, M., {Ros}, E., {Lobanov}, A.~P., {Falcke}, H., \& {Zensus}, J.~A.
  2004, \aap, 426, 481, \dodoi{10.1051/0004-6361:20041051}

\bibitem[{{Kardashev} {et~al.}(2013){Kardashev}, {Khartov}, {Abramov},
  {Avdeev}, {Alakoz}, {Aleksandrov}, {Ananthakrishnan}, {Andreyanov},
  {Andrianov}, {Antonov}, {Artyukhov}, {Arkhipov}, {Baan}, {Babakin},
  {Babyshkin}, {Bartel'}, {Belousov}, {Belyaev}, {Berulis}, {Burke},
  {Biryukov}, {Bubnov}, {Burgin}, {Busca}, {Bykadorov}, {Bychkova},
  {Vasil'kov}, {Wellington}, {Vinogradov}, {Wietfeldt}, {Voitsik},
  {Gvamichava}, {Girin}, {Gurvits}, {Dagkesamanskii}, {D'Addario},
  {Giovannini}, {Jauncey}, {Dewdney}, {D'yakov}, {Zharov}, {Zhuravlev},
  {Zaslavskii}, {Zakhvatkin}, {Zinov'ev}, {Ilinen}, {Ipatov}, {Kanevskii},
  {Knorin}, {Casse}, {Kellermann}, {Kovalev}, {Kovalev}, {Kovalenko}, {Kogan},
  {Komaev}, {Konovalenko}, {Kopelyanskii}, {Korneev}, {Kostenko}, {Kotik},
  {Kreisman}, {Kukushkin}, {Kulishenko}, {Cooper}, {Kut'kin}, {Cannon},
  {Larionov}, {Lisakov}, {Litvinenko}, {Likhachev}, {Likhacheva}, {Lobanov},
  {Logvinenko}, {Langston}, {McCracken}, {Medvedev}, {Melekhin}, {Menderov},
  {Murphy}, {Mizyakina}, {Mozgovoi}, {Nikolaev}, {Novikov}, {Novikov},
  {Oreshko}, {Pavlenko}, {Pashchenko}, {Ponomarev}, {Popov}, {Pravin-Kumar},
  {Preston}, {Pyshnov}, {Rakhimov}, {Rozhkov}, {Romney}, {Rocha}, {Rudakov},
  {R{\"a}is{\"a}nen}, {Sazankov}, {Sakharov}, {Semenov}, {Serebrennikov},
  {Schilizzi}, {Skulachev}, {Slysh}, {Smirnov}, {Smith}, {Soglasnov},
  {Sokolovskii}, {Sondaar}, {Stepan'yants}, {Turygin}, {Turygin}, {Tuchin},
  {Urpo}, {Fedorchuk}, {Finkel'shtein}, {Fomalont}, {Fejes}, {Fomina},
  {Khapin}, {Tsarevskii}, {Zensus}, {Chuprikov}, {Shatskaya}, {Shapirovskaya},
  {Sheikhet}, {Shirshakov}, {Schmidt}, {Shnyreva}, {Shpilevskii}, {Ekers}, \&
  {Yakimov}}]{2013ARep...57..153K}
{Kardashev}, N.~S., {Khartov}, V.~V., {Abramov}, V.~V., {et~al.} 2013,
  Astronomy Reports, 57, 153, \dodoi{10.1134/S1063772913030025}

\bibitem[{{Kardashev} {et~al.}(2017){Kardashev}, {Alakoz}, {Andrianov},
  {Artyukhov}, {Baan}, {Babyshkin}, {Bartel}, {Bayandina}, {Val'tts},
  {Voitsik}, {Vorobyov}, {Gwinn}, {Gomez}, {Giovannini}, {Jauncey}, {Johnson},
  {Imai}, {Kovalev}, {Kurtz}, {Lisakov}, {Lobanov}, {Molodtsov}, {Novikov},
  {Pogodin}, {Popov}, {Privesenzev}, {Rudnitski}, {Rudnitski}, {Savolainen},
  {Smirnova}, {Sobolev}, {Soglasnov}, {Sokolovsky}, {Filippova}, {Khartov},
  {Churikova}, {Shirshakov}, {Shishov}, \& {Edwards}}]{2017SoSyR..51..535K}
{Kardashev}, N.~S., {Alakoz}, A.~V., {Andrianov}, A.~S., {et~al.} 2017, Solar
  System Research, 51, 535, \dodoi{10.1134/S0038094617070085}

\bibitem[{{Kellermann} \& {Pauliny-Toth}(1969)}]{1969ApJ...155L..71K}
{Kellermann}, K.~I., \& {Pauliny-Toth}, I.~I.~K. 1969, \apjl, 155, L71,
  \dodoi{10.1086/180305}

\bibitem[{{Kovalev} {et~al.}(2019){Kovalev}, {Pushkarev}, {Nokhrina}, {Plavin},
  {Beskin}, {Chernoglazov}, {Lister}, \& {Savolainen}}]{2019arXiv190701485K}
{Kovalev}, Y.~Y., {Pushkarev}, A.~B., {Nokhrina}, E.~E., {et~al.} 2019, arXiv
  e-prints, arXiv:1907.01485.
\newblock \doarXiv{1907.01485}

\bibitem[{{Kovalev} {et~al.}(2005){Kovalev}, {Kellermann}, {Lister}, {Homan},
  {Vermeulen}, {Cohen}, {Ros}, {Kadler}, {Lobanov}, {Zensus}, {Kardashev},
  {Gurvits}, {Aller}, \& {Aller}}]{2005AJ....130.2473K}
{Kovalev}, Y.~Y., {Kellermann}, K.~I., {Lister}, M.~L., {et~al.} 2005, \aj,
  130, 2473, \dodoi{10.1086/497430}

\bibitem[{{Kovalev} {et~al.}(2016){Kovalev}, {Kardashev}, {Kellermann},
  {Lobanov}, {Johnson}, {Gurvits}, {Voitsik}, {Zensus}, {Anderson}, {Bach},
  {Jauncey}, {Ghigo}, {Ghosh}, {Kraus}, {Kovalev}, {Lisakov}, {Petrov},
  {Romney}, {Salter}, \& {Sokolovsky}}]{2016ApJ...820L...9K}
{Kovalev}, Y.~Y., {Kardashev}, N.~S., {Kellermann}, K.~I., {et~al.} 2016,
  \apjl, 820, L9, \dodoi{10.3847/2041-8205/820/1/L9}

\bibitem[{{Kuehr} {et~al.}(1981){Kuehr}, {Witzel}, {Pauliny-Toth}, \&
  {Nauber}}]{1981AAS...45..367K}
{Kuehr}, H., {Witzel}, A., {Pauliny-Toth}, I.~I.~K., \& {Nauber}, U. 1981,
  \aaps, 45, 367

\bibitem[{{L{\"a}hteenm{\"a}ki} {et~al.}(1999){L{\"a}hteenm{\"a}ki},
  {Valtaoja}, \& {Wiik}}]{1999ApJ...511..112L}
{L{\"a}hteenm{\"a}ki}, A., {Valtaoja}, E., \& {Wiik}, K. 1999, \apj, 511, 112,
  \dodoi{10.1086/306649}

\bibitem[{{Lee} {et~al.}(2016){Lee}, {Lee}, {Kang}, {Byun}, \&
  {Kim}}]{2016AA...592L..10L}
{Lee}, J.~W., {Lee}, S.-S., {Kang}, S., {Byun}, D.-Y., \& {Kim}, S.~S. 2016,
  \aap, 592, L10, \dodoi{10.1051/0004-6361/201629212}

\bibitem[{{Liao} {et~al.}(2014){Liao}, {Bai}, {Liu}, {Weng}, {Chen}, \&
  {Li}}]{2014ApJ...783...83L}
{Liao}, N.~H., {Bai}, J.~M., {Liu}, H.~T., {et~al.} 2014, \apj, 783, 83,
  \dodoi{10.1088/0004-637X/783/2/83}

\bibitem[{{Lind} \& {Blandford}(1985)}]{1985ApJ...295..358L}
{Lind}, K.~R., \& {Blandford}, R.~D. 1985, \apj, 295, 358,
  \dodoi{10.1086/163380}

\bibitem[{{Lipunov} {et~al.}(2010){Lipunov}, {Kornilov}, {Gorbovskoy},
  {Shatskij}, {Kuvshinov}, {Tyurina}, {Belinski}, {Krylov}, {Balanutsa},
  {Chazov}, {Kuznetsov}, {Kortunov}, {Sankovich}, {Tlatov}, {Parkhomenko},
  {Krushinsky}, {Zalozhnyh}, {Popov}, {Kopytova}, {Ivanov}, {Yazev}, \&
  {Yurkov}}]{2010AdAst2010E..30L}
{Lipunov}, V., {Kornilov}, V., {Gorbovskoy}, E., {et~al.} 2010, Advances in
  Astronomy, 2010, 349171, \dodoi{10.1155/2010/349171}

\bibitem[{{Lister} {et~al.}(2018){Lister}, {Aller}, {Aller}, {Hodge}, {Homan},
  {Kovalev}, {Pushkarev}, \& {Savolainen}}]{2018ApJS..234...12L}
{Lister}, M.~L., {Aller}, M.~F., {Aller}, H.~D., {et~al.} 2018, \apjs, 234, 12,
  \dodoi{10.3847/1538-4365/aa9c44}

\bibitem[{{Lister} {et~al.}(2013){Lister}, {Aller}, {Aller}, {Homan},
  {Kellermann}, {Kovalev}, {Pushkarev}, {Richards}, {Ros}, \&
  {Savolainen}}]{2013AJ....146..120L}
---. 2013, \aj, 146, 120, \dodoi{10.1088/0004-6256/146/5/120}

\bibitem[{{Lister} {et~al.}(2019){Lister}, {Homan}, {Hovatta}, {Kellermann},
  {Kiehlmann}, {Kovalev}, {Max-Moerbeck}, {Pushkarev}, {Readhead}, {Ros}, \&
  {Savolainen}}]{2019ApJ...874...43L}
{Lister}, M.~L., {Homan}, D.~C., {Hovatta}, T., {et~al.} 2019, \apj, 874, 43,
  \dodoi{10.3847/1538-4357/ab08ee}

\bibitem[{{Liu} {et~al.}(2012){Liu}, {Song}, {Marchili}, {Liu}, {Liu},
  {Krichbaum}, {Fuhrmann}, \& {Zensus}}]{2012AA...543A..78L}
{Liu}, X., {Song}, H.-G., {Marchili}, N., {et~al.} 2012, \aap, 543, A78,
  \dodoi{10.1051/0004-6361/201219367}

\bibitem[{{Lobanov}(2015)}]{2015AA...574A..84L}
{Lobanov}, A. 2015, \aap, 574, A84, \dodoi{10.1051/0004-6361/201425084}

\bibitem[{{Lobanov}(2005)}]{2005astro.ph..3225L}
{Lobanov}, A.~P. 2005, ArXiv e-prints.
\newblock \doarXiv{astro-ph/0503225}

\bibitem[{{Lobanov} {et~al.}(2015){Lobanov}, {G{\'o}mez}, {Bruni}, {Kovalev},
  {Anderson}, {Bach}, {Kraus}, {Zensus}, {Lisakov}, {Sokolovsky}, \&
  {Voytsik}}]{2015AA...583A.100L}
{Lobanov}, A.~P., {G{\'o}mez}, J.~L., {Bruni}, G., {et~al.} 2015, \aap, 583,
  A100, \dodoi{10.1051/0004-6361/201526335}

\bibitem[{{MAGIC Collaboration} {et~al.}(2018){MAGIC Collaboration}, {Ahnen},
  {Ansoldi}, {Antonelli}, {Arcaro}, {Baack}, {Babi{\'c}}, {Banerjee},
  {Bangale}, {Barres de Almeida}, {Barrio}, {Becerra Gonz{\'a}lez}, {Bednarek},
  {Bernardini}, {Ch Berse}, {Berti}, {Bhattacharyya}, {Biland}, {Blanch},
  {Bonnoli}, {Carosi}, {Carosi}, {Ceribella}, {Chatterjee}, {Colak}, {Colin},
  {Colombo}, {Contreras}, {Cortina}, {Covino}, {Cumani}, {da Vela}, {Dazzi},
  {de Angelis}, {de Lotto}, {Delfino}, {Delgado}, {di Pierro},
  {Dom{\'{\i}}nguez}, {Dominis Prester}, {Dorner}, {Doro}, {Einecke},
  {Elsaesser}, {Fallah Ramazani}, {Fern{\'a}ndez-Barral}, {Fidalgo}, {Fonseca},
  {Font}, {Fruck}, {Galindo}, {Gallozzi}, {Garc{\'{\i}}a L{\'o}pez},
  {Garczarczyk}, {Gaug}, {Giammaria}, {Godinovi{\'c}}, {Gora}, {Guberman},
  {Hadasch}, {Hahn}, {Hassan}, {Hayashida}, {Herrera}, {Hose}, {Hrupec},
  {Ishio}, {Konno}, {Kubo}, {Kushida}, {Kuve{\v z}di{\'c}}, {Lelas},
  {Lindfors}, {Lombardi}, {Longo}, {L{\'o}pez}, {Maggio}, {Majumdar},
  {Makariev}, {Maneva}, {Manganaro}, {Mannheim}, {Maraschi}, {Mariotti},
  {Mart{\'{\i}}nez}, {Masuda}, {Mazin}, {Mielke}, {Minev}, {Miranda},
  {Mirzoyan}, {Moralejo}, {Moreno}, {Moretti}, {Nagayoshi}, {Neustroev},
  {Niedzwiecki}, {Nievas Rosillo}, {Nigro}, {Nilsson}, {Ninci}, {Nishijima},
  {Noda}, {Nogu{\'e}s}, {Paiano}, {Palacio}, {Paneque}, {Paoletti}, {Paredes},
  {Pedaletti}, {Peresano}, {Persic}, {Prada Moroni}, {Prandini}, {Puljak},
  {Garcia}, {Reichardt}, {Rhode}, {Rib{\'o}}, {Rico}, {Righi}, {Rugliancich},
  {Saito}, {Satalecka}, {Schweizer}, {Sitarek}, {{\v S}nidari{\'c}},
  {Sobczynska}, {Stamerra}, {Strzys}, {Suri{\'c}}, {Takahashi}, {Takalo},
  {Tavecchio}, {Temnikov}, {Terzi{\'c}}, {Teshima}, {Torres-Alb{\`a}},
  {Treves}, {Tsujimoto}, {Vanzo}, {Vazquez Acosta}, {Vovk}, {Ward}, {Will},
  {Zari{\'c}}, {Fermi-Lat Collaboration}, {Bastieri}, {Gasparrini}, {Lott},
  {Rani}, {Thompson}, {MWL Collaborators}, {Agudo}, {Angelakis}, {Borman},
  {Casadio}, {Grishina}, {Gurwell}, {Hovatta}, {Itoh}, {J{\"a}rvel{\"a}},
  {Jermak}, {Jorstad}, {Kopatskaya}, {Kraus}, {Krichbaum}, {Kuin},
  {L{\"a}hteenm{\"a}ki}, {Larionov}, {Larionova}, {Lien}, {Madejski},
  {Marscher}, {Myserlis}, {Max-Moerbeck}, {Molina}, {Morozova}, {Nalewajko},
  {Pearson}, {Ramakrishnan}, {Readhead}, {Reeves}, {Savchenko}, {Steele},
  {Tornikoski}, {Troitskaya}, {Troitsky}, {Vasilyev}, \&
  {Zensus}}]{2018AA...619A..45M}
{MAGIC Collaboration}, {Ahnen}, M.~L., {Ansoldi}, S., {et~al.} 2018, \aap, 619,
  A45, \dodoi{10.1051/0004-6361/201832677}

\bibitem[{{Mart{\'\i}-Vidal} {et~al.}(2012){Mart{\'\i}-Vidal},
  {P{\'e}rez-Torres}, \& {Lobanov}}]{2012A&A...541A.135M}
{Mart{\'\i}-Vidal}, I., {P{\'e}rez-Torres}, M.~A., \& {Lobanov}, A.~P. 2012,
  \aap, 541, A135, \dodoi{10.1051/0004-6361/201118334}

\bibitem[{{Nair} {et~al.}(2019){Nair}, {Lobanov}, {Krichbaum}, {Ros}, {Zensus},
  {Kovalev}, {Lee}, {Mertens}, {Hagiwara}, {Bremer}, {Lindqvist}, \& {de
  Vicente}}]{2019AA...622A..92N}
{Nair}, D.~G., {Lobanov}, A.~P., {Krichbaum}, T.~P., {et~al.} 2019, \aap, 622,
  A92, \dodoi{10.1051/0004-6361/201833122}

\bibitem[{{Nilsson} {et~al.}(2008){Nilsson}, {Pursimo}, {Sillanp{\"a}{\"a}},
  {Takalo}, \& {Lindfors}}]{2008A&A...487L..29N}
{Nilsson}, K., {Pursimo}, T., {Sillanp{\"a}{\"a}}, A., {Takalo}, L.~O., \&
  {Lindfors}, E. 2008, \aap, 487, L29, \dodoi{10.1051/0004-6361:200810310}

\bibitem[{{Ostorero} {et~al.}(2006){Ostorero}, {Wagner}, {Gracia}, {Ferrero},
  {Krichbaum}, {Britzen}, {Witzel}, {Nilsson}, {Villata}, {Bach}, {Barnaby},
  {Bernhart}, {Carini}, {Chen}, {Chen}, {Ciprini}, {Crapanzano}, {Doroshenko},
  {Efimova}, {Emmanoulopoulos}, {Fuhrmann}, {Gabanyi}, {Giltinan},
  {Hagen-Thorn}, {Hauser}, {Heidt}, {Hojaev}, {Hovatta}, {Hroch}, {Ibrahimov},
  {Impellizzeri}, {Ivanidze}, {Kachel}, {Kraus}, {Kurtanidze},
  {L{\"a}hteenm{\"a}ki}, {Lanteri}, {Larionov}, {Lin}, {Lindfors}, {Munz},
  {Nikolashvili}, {Nucciarelli}, {O'Connor}, {Ohlert}, {Pasanen}, {Pullen},
  {Raiteri}, {Rector}, {Robb}, {Sigua}, {Sillanp{\"a}{\"a}}, {Sixtova},
  {Smith}, {Strub}, {Takahashi}, {Takalo}, {Tapken}, {Tartar}, {Tornikoski},
  {Tosti}, {Tr{\"o}ller}, {Walters}, {Wilking}, {Wills}, {Agudo}, {Aller},
  {Aller}, {Angelakis}, {Klare}, {K{\"o}rding}, {Strom}, {Ter{\"a}sranta},
  {Ungerechts}, \& {Vila-Vilar{\'o}}}]{2006AA...451..797O}
{Ostorero}, L., {Wagner}, S.~J., {Gracia}, J., {et~al.} 2006, \aap, 451, 797,
  \dodoi{10.1051/0004-6361:20054075}

\bibitem[{{Pashchenko} {et~al.}(2015){Pashchenko}, {Kovalev}, \&
  {Voitsik}}]{2015CosRe..53..199P}
{Pashchenko}, I.~N., {Kovalev}, Y.~Y., \& {Voitsik}, P.~A. 2015, Cosmic
  Research, 53, 199, \dodoi{10.1134/S0010952515030053}

\bibitem[{{Pearson} {et~al.}(1994){Pearson}, {Shepherd}, {Taylor}, \&
  {Myers}}]{1994AAS...185.0808P}
{Pearson}, T.~J., {Shepherd}, M.~C., {Taylor}, G.~B., \& {Myers}, S.~T. 1994,
  Bulletin of the American Astronomical Society, 26, 1318

\bibitem[{{Planck Collaboration} {et~al.}(2016){Planck Collaboration}, {Ade},
  {Aghanim}, {Arnaud}, {Ashdown}, {Aumont}, {Baccigalupi}, {Banday},
  {Barreiro}, {Bartlett}, \& et~al.}]{2016AA...594A..13P}
{Planck Collaboration}, {Ade}, P.~A.~R., {Aghanim}, N., {et~al.} 2016, \aap,
  594, A13, \dodoi{10.1051/0004-6361/201525830}

\bibitem[{{Pushkarev} {et~al.}(2017){Pushkarev}, {Kovalev}, {Lister}, \&
  {Savolainen}}]{2017MNRAS.468.4992P}
{Pushkarev}, A.~B., {Kovalev}, Y.~Y., {Lister}, M.~L., \& {Savolainen}, T.
  2017, \mnras, 468, 4992, \dodoi{10.1093/mnras/stx854}

\bibitem[{{Quirrenbach} {et~al.}(1989){Quirrenbach}, {Witzel}, {Krichbaum},
  {Hummel}, \& {Alberdi}}]{0716Nature}
{Quirrenbach}, A., {Witzel}, A., {Krichbaum}, T., {Hummel}, C.~A., \&
  {Alberdi}, A. 1989, \nat, 337, 442, \dodoi{10.1038/337442a0}

\bibitem[{{Rani} {et~al.}(2010){Rani}, {Gupta}, {Joshi}, {Ganesh}, \&
  {Wiita}}]{2010ApJ...719L.153R}
{Rani}, B., {Gupta}, A.~C., {Joshi}, U.~C., {Ganesh}, S., \& {Wiita}, P.~J.
  2010, \apjl, 719, L153, \dodoi{10.1088/2041-8205/719/2/L153}

\bibitem[{{Rani} {et~al.}(2015){Rani}, {Krichbaum}, {Marscher}, {Hodgson},
  {Fuhrmann}, {Angelakis}, {Britzen}, \& {Zensus}}]{2015AA...578A.123R}
{Rani}, B., {Krichbaum}, T.~P., {Marscher}, A.~P., {et~al.} 2015, \aap, 578,
  A123, \dodoi{10.1051/0004-6361/201525608}

\bibitem[{{Rastorgueva} {et~al.}(2009){Rastorgueva}, {Wiik}, {Savolainen},
  {Takalo}, {Valtaoja}, {Vetukhnovskaya}, \& {Sokolovsky}}]{2009AA...494L...5R}
{Rastorgueva}, E.~A., {Wiik}, K., {Savolainen}, T., {et~al.} 2009, \aap, 494,
  L5, \dodoi{10.1051/0004-6361:200811425}

\bibitem[{{Readhead}(1994)}]{readhead_94}
{Readhead}, A.~C.~S. 1994, \apj, 426, 51, \dodoi{10.1086/174038}

\bibitem[{{Readhead} {et~al.}(1988){Readhead}, {Nakajima}, {Pearson},
  {Neugebauer}, {Oke}, \& {Sargent}}]{1988AJ.....95.1278R}
{Readhead}, A.~C.~S., {Nakajima}, T.~S., {Pearson}, T.~J., {et~al.} 1988, \aj,
  95, 1278, \dodoi{10.1086/114724}

\bibitem[{{Rickett} {et~al.}(1984){Rickett}, {Coles}, \&
  {Bourgois}}]{1984AA...134..390R}
{Rickett}, B.~J., {Coles}, W.~A., \& {Bourgois}, G. 1984, \aap, 134, 390

\bibitem[{{Rogers} {et~al.}(1974){Rogers}, {Hinteregger}, {Whitney},
  {Counselman}, {Shapiro}, {Wittels}, {Klemperer}, {Warnock}, {Clark}, \&
  {Hutton}}]{1974ApJ...193..293R}
{Rogers}, A.~E.~E., {Hinteregger}, H.~F., {Whitney}, A.~R., {et~al.} 1974,
  \apj, 193, 293, \dodoi{10.1086/153162}

\bibitem[{{Savolainen} {et~al.}(2006){Savolainen}, {Wiik}, {Valtaoja},
  {Kadler}, {Ros}, {Tornikoski}, {Aller}, \& {Aller}}]{2006ApJ...647..172S}
{Savolainen}, T., {Wiik}, K., {Valtaoja}, E., {et~al.} 2006, \apj, 647, 172,
  \dodoi{10.1086/505259}

\bibitem[{{Shepherd}(1997)}]{1997ASPC..125...77S}
{Shepherd}, M.~C. 1997, in Astronomical Society of the Pacific Conference
  Series, Vol. 125, Astronomical Data Analysis Software and Systems VI, ed.
  G.~{Hunt} \& H.~{Payne}, 77

\bibitem[{{Shepherd} {et~al.}(1994){Shepherd}, {Pearson}, \&
  {Taylor}}]{1994BAAS...26..987S}
{Shepherd}, M.~C., {Pearson}, T.~J., \& {Taylor}, G.~B. 1994, in Bulletin of
  the American Astronomical Society, Vol.~26, Bulletin of the American
  Astronomical Society, 987--989

\bibitem[{{Teraesranta} {et~al.}(1998){Teraesranta}, {Tornikoski}, {Mujunen},
  {Karlamaa}, {Valtonen}, {Henelius}, {Urpo}, {Lainela}, {Pursimo}, {Nilsson},
  {Wiren}, {Laehteenmaeki}, {Korpi}, {Rekola}, {Heinaemaeki}, {Hanski},
  {Nurmi}, {Kokkonen}, {Keinaenen}, {Joutsamo}, {Oksanen}, {Pietilae},
  {Valtaoja}, {Valtonen}, \& {Koenoenen}}]{1998AAS..132..305T}
{Teraesranta}, H., {Tornikoski}, M., {Mujunen}, A., {et~al.} 1998, \aaps, 132,
  305, \dodoi{10.1051/aas:1998297}

\bibitem[{{Valtaoja} {et~al.}(1999){Valtaoja}, {L{\"a}hteenm{\"a}ki},
  {Ter{\"a}sranta}, \& {Lainela}}]{1999ApJS..120...95V}
{Valtaoja}, E., {L{\"a}hteenm{\"a}ki}, A., {Ter{\"a}sranta}, H., \& {Lainela},
  M. 1999, \apjs, 120, 95, \dodoi{10.1086/313170}

\bibitem[{{Wagner} {et~al.}(1990){Wagner}, {Sanchez-Pons}, {Quirrenbach}, \&
  {Witzel}}]{1990AA...235L...1W}
{Wagner}, S., {Sanchez-Pons}, F., {Quirrenbach}, A., \& {Witzel}, A. 1990,
  \aap, 235, L1

\bibitem[{{Wagner} \& {Witzel}(1995)}]{1995ARAA..33..163W}
{Wagner}, S.~J., \& {Witzel}, A. 1995, \araa, 33, 163,
  \dodoi{10.1146/annurev.aa.33.090195.001115}

\bibitem[{{Wagner} {et~al.}(1996){Wagner}, {Witzel}, {Heidt}, {Krichbaum},
  {Qian}, {Quirrenbach}, {Wegner}, {Aller}, {Aller}, {Anton}, {Appenzeller},
  {Eckart}, {Kraus}, {Naundorf}, {Kneer}, {Steffen}, \&
  {Zensus}}]{1996AJ....111.2187W}
{Wagner}, S.~J., {Witzel}, A., {Heidt}, J., {et~al.} 1996, \aj, 111, 2187,
  \dodoi{10.1086/117954}

\bibitem[{{Walker} {et~al.}(2000){Walker}, {Dhawan}, {Romney}, {Kellermann}, \&
  {Vermeulen}}]{2000ApJ...530..233W}
{Walker}, R.~C., {Dhawan}, V., {Romney}, J.~D., {Kellermann}, K.~I., \&
  {Vermeulen}, R.~C. 2000, \apj, 530, 233, \dodoi{10.1086/308372}

\end{thebibliography}
\footnotesize

\footnotesize
\begin{longrotatetable}
\begin{deluxetable}{cccccccccccc}
\tablecaption{Results of decomposition of 37\,GHz light curve into individual flares and model fitted results for the core component at 43\,GHz 
\label{tab:decomp}}
\tablecolumns{12}
\tablenum{3}
\tablewidth{0pt}
\tablehead{
\colhead{Flare} & \colhead{$S_\mathrm{max}$} & \colhead{t$_\mathrm{max}^{a}$} & \colhead{$\tau_\mathrm{var}$} & \colhead{log$_{10}$($T_\mathrm{b,var}$)} & \colhead{$\delta_\mathrm{var}$} & \colhead{$S_\mathrm{core}$} & \colhead{BU epoch} & \colhead{Size} & \colhead{log$_{10}$($T_\mathrm{b,rf}$)} & \colhead{log$_{10}$($T_\mathrm{b,int}$)} & \colhead{$\delta$}\\
\colhead{} & \colhead{(Jy)} & \colhead{(y)} & \colhead{(y)}& \colhead{(K)} & \colhead{} & \colhead{(Jy)} & \colhead{(y)}& \colhead{($\mu$as)} & \colhead{(K)} & \colhead{(K)} & \colhead{}
}
\startdata \footnotesize
1 & 1.11$\pm$0.06  & 2012.14 & 0.203$\pm$0.055 & 12.58$\pm$0.09 & 4.2  &\nodata & \nodata & \nodata & \nodata & \nodata & \nodata \\ 
2 & 1.88$\pm$0.14  & 2012.24 & 0.036$\pm$0.007 & 14.30$\pm$0.16 & 15.9 & \nodata & \nodata & \nodata & \nodata & \nodata & \nodata \\ 
3 & 1.35$\pm$0.25  & 2012.55 & 0.045$\pm$0.016 & 13.98$\pm$0.52 & 12.4 & \nodata & \nodata & \nodata & \nodata & \nodata & \nodata \\ 
4 & 2.78$\pm$0.34  & 2012.71 & 0.026$\pm$0.005 & 14.75$\pm$0.21 & 22.5 & \nodata & \nodata & \nodata & \nodata & \nodata & \nodata \\ 
5 & 4.60$\pm$0.22  & 2012.85 & 0.081$\pm$0.020 & 13.99$\pm$0.17 & 12.5 & 3.46$\pm$0.18 & 2012.83 & 47.1$\pm$0.4 & 12.13$\pm$0.02 & 11.20$\pm$0.09 & 8.5$\pm$0.2\\
6 & 2.91$\pm$0.32  & 2012.95 & 0.028$\pm$0.014 & 14.70$\pm$0.44 & 21.6 & 2.87$\pm$0.14$^{b}$ & 2012.94 & 21.8$\pm$0.9 &12.72$\pm$0.04 & 11.73$\pm$0.35 & 9.8$\pm$0.5\\
7 & 5.37$\pm$0.27  & 2013.02 & 0.022$\pm$0.003 & 15.18$\pm$0.14 & 31.2 & 3.35$\pm$0.18 & 2013.04 & 20.1$\pm$1.3 & 12.86$\pm$0.06 & 11.69$\pm$0.12 & 14.5$\pm$1.0 \\
8 & 3.63$\pm$0.50  & 2013.11 & 0.025$\pm$0.006 & 14.91$\pm$0.32 & 25.3 & \nodata & \nodata & \nodata & \nodata & \nodata & \nodata \\ 
9 & 2.02$\pm$0.14& 2013.20& 0.036$\pm$0.004& 14.35$\pm$0.21& 16.5 & \nodata & \nodata & \nodata & \nodata & \nodata & \nodata \\                                   
10& 4.92$\pm$0.12& 2013.53& 0.067$\pm$0.003& 14.18$\pm$0.04& 14.5 & \nodata & \nodata & \nodata & \nodata & \nodata & \nodata \\ 
11& 2.67$\pm$0.26& 2013.64& 0.003$\pm$0.001& 16.63$\pm$0.51& 94.5 & \nodata & \nodata & \nodata & \nodata & \nodata & \nodata \\ 
12& 2.83$\pm$0.04& 2013.91& 0.348$\pm$0.017& 12.52$\pm$0.04&  4.0 &2.56$\pm$0.13$^{c}$& 2013.87& 34.2$\pm$2.8& 12.28$\pm$0.08& 12.16$\pm$0.12&  1.3$\pm$0.1\\
13& 1.25$\pm$0.08& 2014.34& 0.029$\pm$0.004& 14.31$\pm$0.10& 16.0 &1.99$\pm$0.10&   2014.34& 30.6$\pm$1.2& 12.26$\pm$0.04& 11.24$\pm$0.08& 10.5$\pm$0.5\\
14& 3.78$\pm$0.10& 2014.62& 0.090$\pm$0.005& 13.82$\pm$0.05& 10.9 & \nodata & \nodata & \nodata & \nodata & \nodata & \nodata \\ 
15& 2.77$\pm$0.24& 2014.79& 0.016$\pm$0.002& 15.17$\pm$0.23& 31.0 & \nodata & \nodata & \nodata & \nodata & \nodata & \nodata \\ 
16& 3.91$\pm$0.09& 2014.89& 0.044$\pm$0.002& 14.45$\pm$0.04& 17.8 & \nodata & \nodata & \nodata & \nodata & \nodata & \nodata \\ 
17& 4.38$\pm$0.09& 2015.09& 0.037$\pm$0.002& 14.66$\pm$0.04& 20.9 & \nodata & \nodata & \nodata & \nodata & \nodata & \nodata \\ 
18& 3.35$\pm$0.13& 2015.19& 0.025$\pm$0.002& 14.87$\pm$0.08& 24.6 & \nodata & \nodata & \nodata & \nodata & \nodata & \nodata \\ 
19& 2.28$\pm$0.12& 2015.28& 0.036$\pm$0.004& 14.38$\pm$0.09& 16.9 &2.09$\pm$0.10&   2015.28& 16.54$\pm$1.2& 12.82$\pm$0.07& 12.04$\pm$0.11&  6.0$\pm$0.5\\
20& 1.62$\pm$0.20& 2015.38& 0.013$\pm$0.002& 15.15$\pm$0.51& 30.5 & \nodata & \nodata & \nodata & \nodata & \nodata & \nodata \\ 
21& 3.04$\pm$0.07& 2015.54& 0.124$\pm$0.009& 13.44$\pm$0.06&  8.2 & \nodata & \nodata & \nodata & \nodata & \nodata & \nodata \\ 
22& 1.48$\pm$0.20& 2015.68& 0.016$\pm$0.004& 14.90$\pm$0.23& 25.2 & \nodata & \nodata & \nodata & \nodata & \nodata & \nodata \\ 
23& 2.12$\pm$0.09& 2015.80& 0.121$\pm$0.010& 13.31$\pm$0.04&  7.4 & \nodata & \nodata & \nodata & \nodata & \nodata & \nodata \\ 
24& 1.30$\pm$0.35& 2016.05& 0.012$\pm$0.004& 15.11$\pm$0.12& 29.5 & \nodata & \nodata & \nodata & \nodata & \nodata & \nodata \\ 
25& 3.16$\pm$0.16& 2016.30& 0.032$\pm$0.002& 14.63$\pm$0.06& 20.5 & 3.15$\pm$0.16&   2016.31& 24.1$\pm$0.5& 12.67$\pm$0.03& 11.69$\pm$0.05&  9.5$\pm$0.3\\
26& 2.34$\pm$0.03& 2016.65& 0.435$\pm$0.029& 12.24$\pm$0.05&  3.3 & \nodata & \nodata & \nodata & \nodata & \nodata & \nodata \\ 
27& 1.83$\pm$0.28& 2016.92& 0.024$\pm$0.005& 14.67$\pm$0.15& 21.0 & 2.47$\pm$0.12&   2016.91& 19.5$\pm$0.7& 12.75$\pm$0.04& 11.79$\pm$0.09&  9.1$\pm$0.4\\
\enddata
\tablecomments{Decomposed flares are shown in Fig.~\ref{fig:tb}. See Sect.~\ref{s:orient} for details.
$^{a}$ Typical error is of the order of a day. $^{b}$ Fitted model is not reliable. $^{c}$ The core is modelled by elliptical Gaussian component.}
\end{deluxetable}
\end{longrotatetable}

\end{document}